\pdfoutput=1

\documentclass[journal]{IEEEtran}
\usepackage[numbers,sort&compress]{natbib}

\usepackage{amsmath}
\usepackage{amssymb}

\usepackage{tikz}
\usetikzlibrary{shapes,arrows,positioning,fit}
\usepackage{tabularx}
\usepackage{xcolor}
\definecolor{rasOrange}{rgb}{1,0.7451,0.0039}
\usepackage[nolist,nohyperlinks]{acronym}

\usepackage{bm}
\let\vec\bm

\begin{document}

\title{Geometric Constellation Shaping for Fiber Optic Communication Systems via End-to-end Learning}

\author{Rasmus~T.~Jones,~\IEEEmembership{Student Member,~IEEE,}
        Tobias~A.~Eriksson,
        Metodi~P.~Yankov,~\IEEEmembership{Member,~IEEE,}
        Benjamin~J.~Puttnam,~\IEEEmembership{Member,~IEEE,}
       	Georg~Rademacher,~\IEEEmembership{Member,~IEEE,}
		Ruben~S.~Lu\'is,~\IEEEmembership{Member,~IEEE,}
        and~Darko~Zibar
\thanks{R.T.~Jones and~D.~Zibar are with the Department of Photonics Engineering, Technical University of Denmark, 2800 Kgs. Lyngby, Denmark, e-mail: rajo@fotonik.dtu.dk}
\thanks{M.P.~Yankov is with Fingerprint Cards A/S, 2730 Herlev, Denmark, and with the Department of Photonics Engineering, Technical University of Denmark, 2800 Kgs. Lyngby, Denmark.}
\thanks{T.A.~Eriksson is with the Quantum ICT Advanced Research Center, and B.J.~Puttnam, G.~Rademacher and R.S.~Lu\'is are with the Photonic Network System Laboratory at the National Institute of Information and Communications Technology, 4-2-1 Nukui-Kita Machi, Koganei, Tokyo 184-8795.}
}

\markboth{}%
{}

\maketitle

\begin{abstract}
In this paper, an unsupervised machine learning method for geometric constellation shaping is investigated. By embedding a differentiable fiber channel model within two neural networks, the learning algorithm is optimizing for a geometric constellation shape. The learned constellations yield improved performance to state-of-the-art geometrically shaped constellations, and include an implicit trade-off between amplification noise and nonlinear effects.
Further, the method allows joint optimization of system parameters, such as the optimal launch power, simultaneously with the constellation shape. An experimental demonstration validates the findings.
Improved performances are reported, up to 0.13~bit/4D in simulation and experimentally up to 0.12~bit/4D.
\end{abstract}

\begin{IEEEkeywords}Optical fiber communication, constellation shaping, machine learning, neural networks
\end{IEEEkeywords}

\IEEEpeerreviewmaketitle

\tikzstyle{block} = [rectangle, draw, text centered, minimum width=14em, minimum height=1.5em]
\tikzstyle{line} = [draw, -latex']
\tikzstyle{ppp} = [rectangle, draw, text centered, minimum height=1.5em]

\section{Introduction}
\label{sec:introduction}
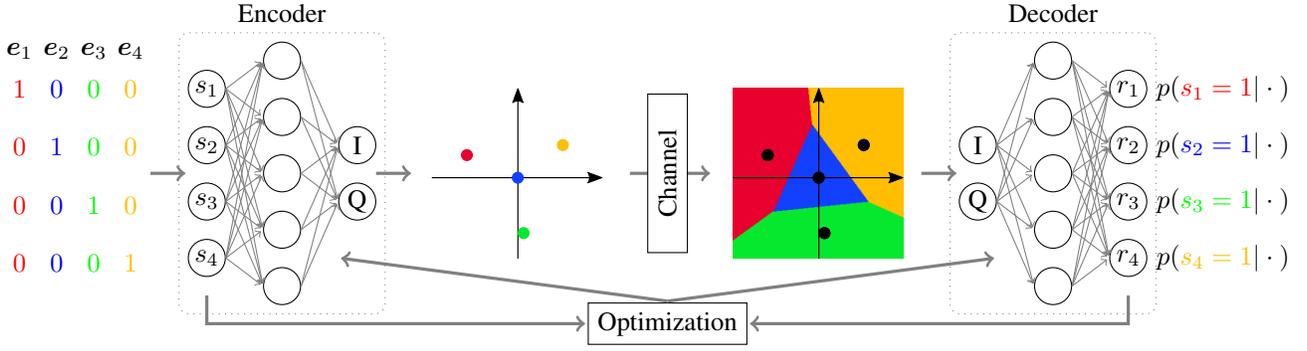
\begin{figure*}[t]
\centering
\begin{tikzpicture}[shorten >=1pt,->,draw=black!50, node distance=\layersep]
	\tikzstyle{block} = [rectangle,draw=black, text centered, minimum width=6em, minimum height=1.5em]
    \tikzstyle{neuron}=[circle,fill=white,draw=black,minimum size=14pt,inner sep=0pt]
    \tikzstyle{annot} = [text width=4em, text centered]
	\tikzstyle{line} = [draw, --]
	\tikzstyle{arrow} = [draw, -latex']
	
	\def\names{{"A","I","Q"}}%
	\def\inputnames{{"A","$s_1$","$s_2$","$s_3$","$s_4$"}}%
	\def\outputnames{{"A","$r_1$","$r_2$","$r_3$","$r_4$"}}%
	\def\outputnamesRight{{"A","$p(s_1=1|{}\cdot{})$","$p(s_2=1|{}\cdot{})$","$p(s_3=1|{}\cdot{})$","$p(s_4=1|{}\cdot{})$"}}%
    
    \foreach \name / \y in {1,...,4}
        \node[neuron] (eI-\name) at (0,-0.75*\y) {\pgfmathparse{\inputnames[\y]}\pgfmathresult};
    \foreach \name / \y in {1,...,5}
        \path[yshift=0.375cm]
            node[neuron] (eH-\name) at (1.00cm,-0.75*\y) {};	
    \foreach \name / \y in {1,...,2}
    	\path[yshift=-0.75cm]
    		node[neuron] (eO-\name) at (2.00cm,-0.75*\y) {\pgfmathparse{\names[\y]}\pgfmathresult};
    		
    \foreach \source in {1,...,4}
        \foreach \dest in {1,...,5}
            \path (eI-\source.east) edge (eH-\dest.west);
    \foreach \source in {1,...,5}
    	\foreach \dest in {1,...,2}
        	\path (eH-\source.east) edge (eO-\dest.west);
	
	\coordinate[left of=eH-3, node distance=1.75cm] (arrow1Start);	
	\coordinate[left of=eH-3, node distance=1.25cm] (arrow1End);
	\draw[very thick] (arrow1Start) -- (arrow1End);
	
	\node[node distance=2.75cm, left of=eH-3,yshift=0.25cm] (vecsIn) {
    $   \begingroup 
			\setlength\arraycolsep{2pt}
			\begin{matrix}
	\vec{e}_1 & \vec{e}_2 & \vec{e}_3 & \vec{e}_4 \\[2pt]
    {\color{red}1} & {\color{blue}0} & {\color{green}0} & {\color{rasOrange}0} \\[8pt]
    {\color{red}0} & {\color{blue}1} & {\color{green}0} & {\color{rasOrange}0} \\[8pt]
    {\color{red}0} & {\color{blue}0} & {\color{green}1} & {\color{rasOrange}0} \\[8pt]
    {\color{red}0} & {\color{blue}0} & {\color{green}0} & {\color{rasOrange}1} \\
			\end{matrix}
		\endgroup
	$
	};
	
	\coordinate[right of=eH-3, node distance=1.25cm] (arrow2Start);	
	\coordinate[right of=eH-3, node distance=1.75cm] (arrow2End);
	\draw[very thick] (arrow2Start) -- (arrow2End);
	
	\node[right of=eH-3,node distance=3.125cm] (points)
		{\includegraphics[width=0.125\linewidth]{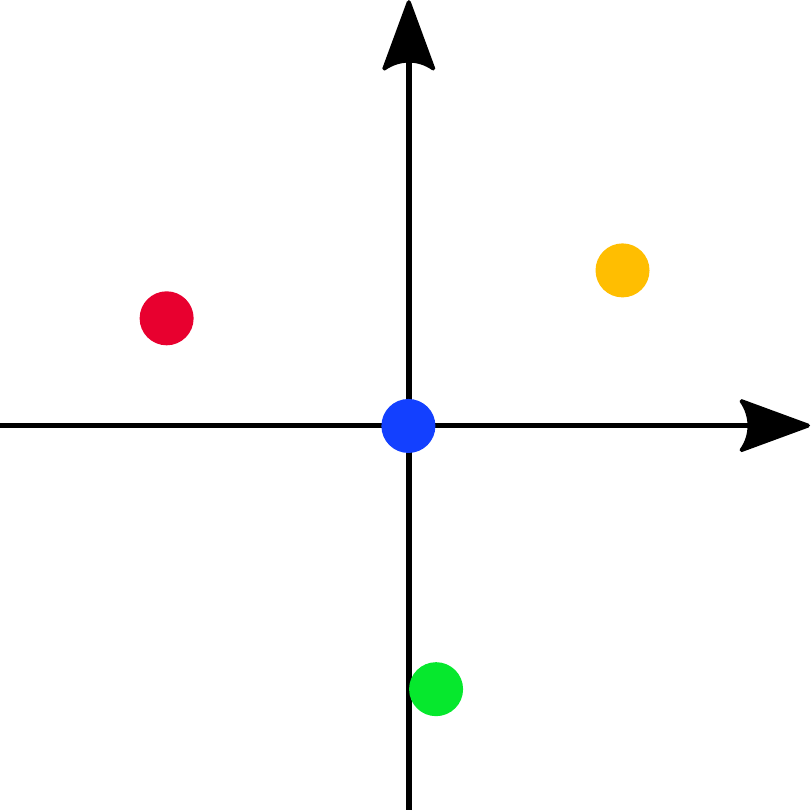}};
		
	\coordinate[right of=points, node distance=1.5cm] (arrow3Start);	
	\coordinate[right of=points, node distance=2.6cm] (arrow3End);
	\draw[very thick] (arrow3Start) -- (arrow3End);
	
	\node [block,right of=points,node distance=2cm,fill=white,rotate=90] (channel) {Channel};
	
	\node[right of=channel,node distance=2cm] (pointsDecision)
		{\includegraphics[width=0.125\linewidth]{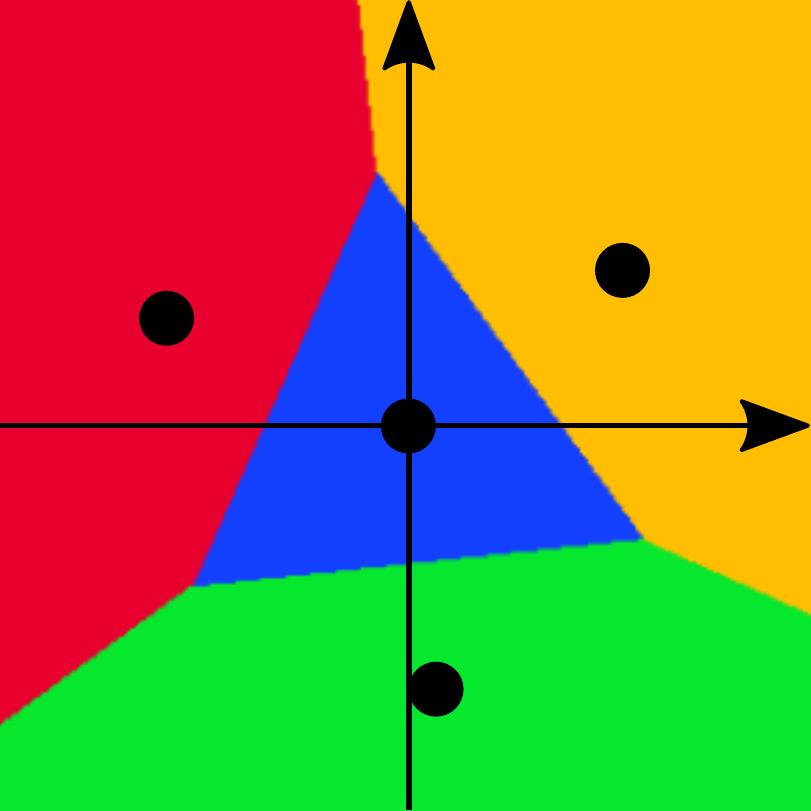}};
		
	\coordinate[right of=pointsDecision, node distance=1.375cm] (arrow4Start);	
	\coordinate[right of=pointsDecision, node distance=1.875cm] (arrow4End);
	\draw[very thick] (arrow4Start) -- (arrow4End);
		
    \foreach \name / \y in {1,...,2}
    	\path[yshift=-0.75cm]
        	node[neuron] (dI-\name) at (10.25cm,-0.75*\y) {\pgfmathparse{\names[\y]}\pgfmathresult};
    \foreach \name / \y in {1,...,5}
        \path[yshift=0.375cm]
            node[neuron] (dH-\name) at (11.25cm,-0.75*\y) {};	
    \node[neuron,label=right:$p({\color{red}s_1=1}|{}\cdot{})$] (dO-1) at (12.25cm,-0.75*1) {$r_1$};
    \node[neuron,label=right:$p({\color{blue}s_2=1}|{}\cdot{})$] (dO-2) at (12.25cm,-0.75*2) {$r_2$};
    \node[neuron,label=right:$p({\color{green}s_3=1}|{}\cdot{})$] (dO-3) at (12.25cm,-0.75*3) {$r_3$};
    \node[neuron,label=right:$p({\color{rasOrange}s_4=1}|{}\cdot{})$] (dO-4) at (12.25cm,-0.75*4) {$r_4$};
    
    \foreach \source in {1,...,2}
        \foreach \dest in {1,...,5}
            \path (dI-\source.east) edge (dH-\dest.west);
    \foreach \source in {1,...,5}
    	\foreach \dest in {1,...,4}
        	\path (dH-\source.east) edge (dO-\dest.west);
	
	\node [block,below of=channel, node distance=2.0cm] (opt) {Optimization};
	\coordinate[below of=eI-4, node distance=0.5cm] (optLeftStart);
	\coordinate[left of=opt, node distance=6.125cm] (optLeft);
	\draw[very thick] (optLeftStart) -- (optLeft) -- (opt);
	
	\coordinate[below of=dO-4, node distance=0.5cm] (optRightStart);
	\coordinate[right of=opt, node distance=6.125cm] (optRight);
	\draw[very thick] (optRightStart) -- (optRight) -- (opt);
	
	\coordinate[above of=opt, node distance=0.3cm] (optTop);
	\coordinate (optTopLeft) at (1.75cm,-3cm);
	\draw[very thick] (optTop) -- (optTopLeft);
	
	\coordinate (optTopRight) at (10.5cm,-3cm);
	\draw[very thick] (optTop) -- (optTopRight);
	
	\node[draw,dotted,rounded corners,fit=(eI-1) (eI-2) (eI-3) (eI-4) (eH-1) (eH-2) (eH-3) (eH-4) (eH-5) (eO-1) (eO-2)] (labelOne) {};
	\node[annot,above of=eH-1, node distance=0.625cm] (il) {Encoder};
	
	\node[draw,dotted,rounded corners,fit=(dO-1) (dO-2) (dO-3) (dO-4) (dH-1) (dH-2) (dH-3) (dH-4) (dH-5) (dI-1) (dI-2)] (labelTwo) {};
	\node[annot,above of=dH-1, node distance=0.625cm] (ol) {Decoder};
	
\end{tikzpicture}
    \caption{Two neural networks, encoder and decoder, wrap a channel model and are optimized end-to-end. Given multiple instances of four different vectors, the encoder learns four constellation symbols which are robust to the channel impairments. Simultaneously, the decoder learns to classify the received symbols such that the initial vectors are reconstructed.}
    \label{fig:autoEncoderBaby}
\end{figure*}
In order to meet future demands in data traffic applications, optical communication systems have to offer higher spectral efficiency~\cite{essiambre2010capacity}. Coherent detection has enabled advanced optical modulation formats that have led to increased data throughput as well as allowing constellation shaping. For the \ac{AWGN} channel, constellation shaping offers gains of up to 1.53~dB in \ac{SNR} through Gaussian shaped constellations~\cite{fehenberger2016probabilistic,peric1998design}. Iterative polar modulation\acused{IPM} (\ac{IPM})-based geometrically shaped constellations were introduced in~\cite{peric1998design} and thereafter applied in optical communications~\cite{djordjevic2010coded,batshon2010iterative}. The iterative optimization method optimizes for a geometric shaped constellation of multiple rings to achieve shaping gain. However, the ultimate shaping gain for the nonlinear fiber channel is unknown. An optimal constellation for the optical channel is jointly robust to amplification noise and signal dependent nonlinear interference~\cite{geller2016shaping,yankovn2016constellation,sillekens2018simple,sillekens2018experimental}.
The signal degrading nonlinearities are dependent on the high-order moments of the transmitted constellation~\cite{dar2013properties}. A shaped constellation with low high-order moments reduces the nonlinear distortions~\cite{dar2014accumulation}.
This is in contradiction with Gaussian shaped constellations robust to amplification noise, which hold comparatively large high-order moments.
Hence, an optimal set of high-order moments exists, which maximizes the available \ac{SNR} at the receiver.
Machine learning is an established technique for learning complex relationships between signals and is rapidly entering the field of optical communications~\cite{chen2006channel,shen2011fiber,khan2016modulation,khan2017machine,zibar2017machine,eriksson2017applying,koike2018fiber,hager2018deep,karanov2018end,li2018achievable,Caballero2018,musumeci2018survey}. In particular, the autoencoder has been used to learn how to communicate in a range of complex systems and channels~\cite{o2017introduction}. It builds upon a differentiable channel model wrapped by two neural networks~\cite{goodfellow2016deep} (encoder and decoder), which are trained jointly by minimizing the reconstruction error, as shown in Fig.~\ref{fig:autoEncoderBaby}. By embedding an optical fiber channel model within an autoencoder, improved performance can be achieved in \ac{IM-DD} systems~\cite{karanov2018end}. Similarly, in this paper we propose to learn geometric constellation shapes for coherent systems, such that channel model characteristics are captured by the autoencoder, leading to constellations jointly robust to amplification noise and fiber nonlinearities.
When the autoencoder is trained on the \ac{GN}-model~\cite{poggiolini2012gn}, the learned constellation is optimized for an \ac{AWGN} channel with an effective \ac{SNR} determined by the launch power, and nonlinear effects are not mitigated. When it is trained on the \ac{NLIN}-model~\cite{dar2013properties,dar2014accumulation} the learned constellation mitigates nonlinear effects by optimizing its high-order moments.
In this work, the performance in terms of \ac{MI} and estimated received \ac{SNR} of the learned constellations is compared to standard \ac{QAM} and \ac{IPM}-based geometrically shaped constellations. The \ac{IPM}-based geometrically shaped constellations optimized under \ac{AWGN} assumption provided in~\cite{peric1998design,djordjevic2010coded,batshon2010iterative} are used in this paper. 
From the simulations and experimental demonstrations, gains of 0.13~bit/4D and 0.12~bit/4D, respectively, were observed with respect to \ac{IPM}-based geometrically shaped constellations in the weakly nonlinear region of transmission around the optimal launch power.
The experimental validation also revealed the challenge of matching the chosen channel model of the training process to the experimental setup.

This paper is an extension of our conference paper~\cite{jones2018deep}, which includes a similar study on a 2000~km (20~spans) simulated transmission link. The extension presented here provides an experimental demonstration, a more rigorous mathematical description of the autoencoder model, a numerical simulation study for a 1000~km (10~spans) link, and a generalization of the method towards jointly learning system parameters and geometric shaped constellations.

The paper is organized as follows. Section~\ref{sec:models} describes the channel models used. Section~\ref{sec:endtoendlearning} describes the autoencoder architecture, its training, and joint optimization of system parameters.
Section~\ref{sec:shapingViaEndToEndLearning} describes the simulation results evaluating the learned constellations with the \ac{NLIN}-model and \ac{SSF} method. The experimental demonstration and results are also presented in this Section. In Section~\ref{sec:discussion} the simulation and experimental results are discussed and thereafter concluded in Section~\ref{sec:conclusion}.
The optical fiber models together with the autoencoder model are available online as Python/TensorFlow programs~\cite{claude2018}.

\section{Fiber Models}
\label{sec:models}
The \ac{NLSE} describes the propagation of light through an optical fiber. The \ac{SSF} method approximates the outcome by solving the \ac{NLSE} through many consecutive numerical simulation steps.
In contrast, models of optical communication systems allow analysis of the performance of such systems whilst avoiding cumbersome simulations using the \ac{SSF} method. In particular, simulations of \ac{WDM} systems and their nonlinear effects are computational expensive via the \ac{SSF} method. Modeling of such systems~\cite{poggiolini2012gn,dar2013properties,carena2014egn} allows more efficient analysis. The \ac{GN}-model~\cite{poggiolini2012gn} assumes statistical independence of the frequency components within all interfering channels and hence models the nonlinear interference as a memoryless \ac{AWGN} term dependent on the launch power per channel.
This assumption is not made in the \ac{NLIN}-model~\cite{dar2013properties} which can therefore include modulation dependent effects and allows more accurate analysis of non-conventional modulation schemes, such as probabilistic and geometric shaped constellations~\cite{fehenberger2016probabilistic}.
The \ac{EGN} model~\cite{carena2014egn} is an extension of the above two models, including additional cross wavelength interactions, which are only significant in densely packed \ac{WDM} systems.
Hence, for the system investigated here, it is sufficient to use the \ac{NLIN}-model, while the \ac{GN}-model allows the study under an \ac{AWGN} channel assumption independent of modulation dependent effects.
The \ac{NLIN}-model describes the effective \ac{SNR} of a fiber optic system as follows:
\begin{equation}
    \text{SNR}_{\text{eff}} = \frac{P_{\text{tx}}}{\sigma^2_{\text{ASE}} + \sigma^2_{\text{NLIN}}(P_{\text{tx}},\mu_4,\mu_6)},
    \label{eq:effSNR}
\end{equation}
where $P_{\text{tx}}$ is the optical launch power, $\sigma^2_{\text{ASE}}$ is the variance of the accumulated \ac{ASE} noise and $\sigma^2_{\text{NLIN}}(\cdot)$ is the variance of the \ac{NLIN}. The parameters $\mu_4$ and $\mu_6$ describe the fourth and sixth order moment of the constellation, respectively. Here, all channels are assumed to have the same optical launch power and draw from the same constellation.
We refer to $\sigma^2_{\text{NLIN}}(\cdot)$ as a function of the optical launch power and moments of the constellation, since these parameters are optimized throughout this work. The \ac{NLIN} further depends on system specific terms, which are estimated through Monte-Carlo integration and are constant when not altering the system.
A per symbol description of the memoryless model follows:
\begin{equation}
    \begin{split}
        y[k] &= c_{\text{NLIN}}(x[k],P_{\text{tx}},\mu_4,\mu_6),\\
             &= x[k] + n_{\text{ASE}}[k] + n_{\text{NLIN}}[k],
    \end{split}
\end{equation}
where $x[k]$ and $y[k]$ are the transmitted and received symbols at time $k$, $c_{\text{NLIN}}(\cdot)$ is the channel model, and $n_{\text{ASE}}[k] \sim N(0,\sigma^2_{\text{ASE}})$ and $n_{\text{NLIN}}[k] \sim N(0,\sigma^2_{\text{NLIN}}(\cdot))$ are Gaussian noise samples with variances $\sigma^2_{\text{ASE}}$ and $\sigma^2_{\text{NLIN}}(\cdot)$, respectively.
Since the \ac{GN}-model is independent of the moments of the constellation, its variance term reduces to $\sigma^2_{\text{GN}}(P_{\text{tx}})$ and its channel model to $c_{\text{GN}}(x[k],P_{\text{tx}})$.
All considered channel models are memoryless. However, the interaction between dispersion and nonlinearities introduces memory.
Models including memory effects are available~\cite{agrell2014capacity,dar2015inter,golani2016correlations,golani2016modeling}, and we note that in combination with temporal machine learning algorithms, e.g. recurrent neural networks~\cite{goodfellow2016deep}, these might yield larger gains. Such study is left for future research.

\section{End-to-end Learning of Communication Systems}
\label{sec:endtoendlearning}
\subsection{Autoencoder Model Architecture}
\begin{figure}[b]
\centering
\begin{tikzpicture}[auto, node distance=1.5cm]
    \coordinate (init);
    \node [block,below of=init] (NNone) {Dense Neural Network};
    \node [block,anchor=north] at (NNone.south) (r2c) {$\mathbb{R}^{2 N} \rightarrow \mathbb{C}^{N}$};
    \node [block,anchor=north] at (r2c.south) (norm) {Normalization};
    
    \node [block,below of=norm] (channel) {Channel};
    
    
    \node [block, below of=channel] (c2r) {$\mathbb{C}^{N} \rightarrow \mathbb{R}^{2 N}$};
    \node [block,anchor=north] at (c2r.south) (NNtwo) {Dense Neural Network};
    \node [block,anchor=north] at (NNtwo.south) (softmax) {Softmax};
    
    \coordinate[below of=softmax] (end);
    
    \draw [-latex'] (init) to node[auto] {$\vec{s} \in S=\{ \vec{e}_i~|~i=1..M \}$} (NNone);
    \path [line] (norm) -- node[auto] {$\vec{x} \in \mathbb{C}^N$} (channel);
    \path [line] (channel) -- node[auto] {$\vec{y} \in \mathbb{C}^N$} (c2r);
    \draw [-latex'] (softmax) to node[auto] {$\vec{r} \in \{\vec{p} \in \mathbb{R}^{M}_{+}| \sum^{M}_{i=1} p_{i} = 1\}$} (end);

    \node[draw,dotted,rounded corners,fit=(NNone) (r2c) (norm)] (labelOne) {};
    \node[right, inner sep=2pt] at (labelOne.east) {$f_{\vec{\theta_f}}(\cdot)$};
    \node[above right, inner sep=2pt] at (labelOne.north west) {Encoder};
    
    \node[draw,dotted,rounded corners,fit=(channel)] (labelTwo) {};
    \node[right, inner sep=2pt] at (labelTwo.east) {$c_{\text{GN/NLIN}}(\cdot)$};
    
    \node[draw,dotted,rounded corners,fit=(c2r) (NNtwo) (softmax)] (labelThree) {};
    \node[right, inner sep=2pt] at (labelThree.east) {$g_{\vec{\theta_g}}(\cdot)$};
    \node[above right, inner sep=2pt] at (labelThree.north west) {Decoder};
\end{tikzpicture}
\caption{End-to-end autoencoder model.}
\label{fig:autoEncoder}
\end{figure}
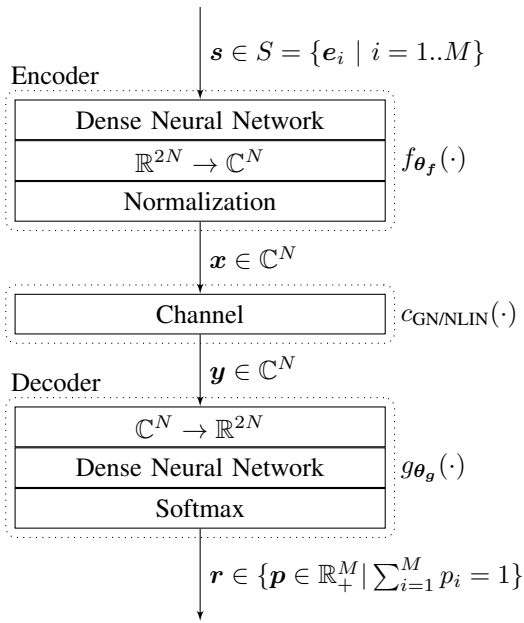
An autoencoder is composed of two parametric functions, an encoder and a decoder, with the goal to reproduce its input vector at the output. The hidden layer in between the encoder and decoder, the latent space, is of lower dimension than the input vector. Thereby, the encoder must learn a meaningful representation of the input vector, which provided to the decoder, holds enough information for replication. By embedding a channel model within such an autoencoder, it is bound to learn a representation robust to the channel impairments. An autoencoder model, with neural networks~\cite{goodfellow2016deep} as encoder and decoder, is depicted in Fig. \ref{fig:autoEncoder} and mathematically described as follows:
t\begin{equation}
\begin{split}
    \vec{x} &= f_{\vec{\theta_f}}(\vec{s}), \\
    \vec{y} &= c_{\text{GN/NLIN}}(\vec{x}), \\
    \vec{r} &= g_{\vec{\theta_g}}(\vec{y}), \\
    \end{split}
\end{equation}
where $f_{\vec{\theta_f}}(\cdot)$ is the encoder, $g_{\vec{\theta_g}}(\cdot)$ the decoder and $c_{\text{GN/NLIN}}(\cdot)$ the channel model. The goal is to reproduce the input $\vec{s}$ at the output $\vec{r}$ through the latent variable $\vec{x}$ (and its impaired version $\vec{y}$). The trainable variables (weights and biases) of the encoder and decoder neural networks are represented by $\vec{\theta_f}$ and $\vec{\theta_g}$, respectively. The parameter vector $\vec{\theta} = \{\vec{\theta_f},\vec{\theta_g}\}$ holds all trainable variables.
The encoder is optimizing the location of the constellation points, at the same time as the decoder learns decision boundaries in between the impaired symbols.
In order to learn a geometrical constellation shape, the structure of the autoencoder must be aligned to the properties of the desired constellation. The dimension of input and output space is equal to the order of the constellation, and the dimension of the latent space is equal to the dimension of the constellation.
A constellation of order $M$ is trained with so called one-hot encoded vectors~$\vec{s} \in S=\{ \vec{e}_i~|~i=1..M \}$, where $\vec{e}_i$ is equal 1 at row $i$ and else 0. The decoder, concluded with a softmax function~\cite{goodfellow2016deep}, yields a probability vector $\vec{r} \in \{\vec{p} \in \mathbb{R}^{M}_{+}| \sum^{M}_{i=1} p_{i} = 1\}$.
A constellation of $N$~complex dimensions is learned by choosing the output of the encoder network and input of the decoder network to hold $2 N$~real dimensions.
The normalization before the channel poses an average power constraint on the learned constellation.

\subsection{Autoencoder Model Training}\label{subsec:AutoencoderModelTraining}
The autoencoder model parameters $\vec{\theta}$ are trained by minimizing the cross-entropy loss~\cite{goodfellow2016deep}:
\begin{equation}
	L(\vec{\theta}) = \underbrace{ \frac{1}{K} \sum^K_{k=1} \overbrace{ \left[ - \sum^M_{i=1} s^{(k)}_{i} \log(r^{(k)}_{i}) \right] }^{\text{Cross-entropy}} }_{\text{Expectation}},
	\label{eq:loss}
\end{equation}
where $K$ is the training batch size. Since $\vec{s}$ only holds a single non-zero value, the inner summation over $M$ requires only one evaluation. The average of the cross-entropy over all samples is computed (\ref{eq:loss}) and through them an estimate of the gradient w.r.t. the parameters of the model. The gradient is used to update the parameters such that the loss is minimized~\cite{o2017introduction}:

\begin{figure}[b]
	\centering
	\includegraphics[width=\linewidth]{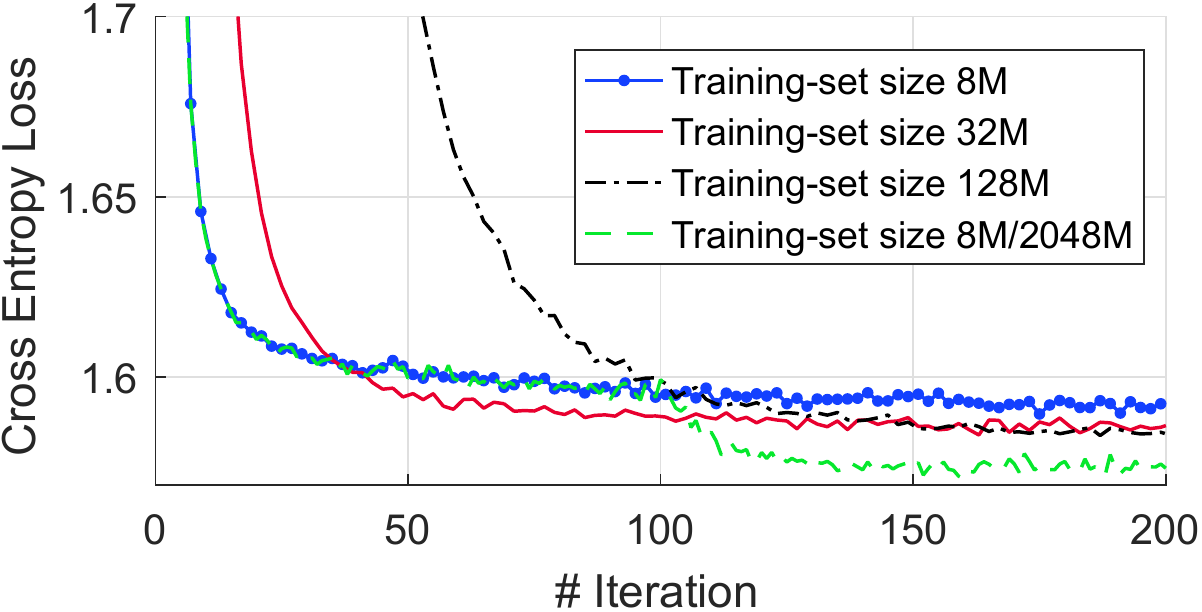}
    \caption{Convergence of cross entropy loss for different training batch sizes. The green/dashed curve is trained with a training batch size of $8 M$ until iteration 100, from where it is increased to $2048 M$. This example learns a constellation with the \ac{NLIN}-model for $M$=64.}
    \label{fig:bsEntropy}
\end{figure}

\begin{equation}
	\vec{\theta}^{(j+1)} = \vec{\theta}^{(j)} - \eta \nabla_{\theta} \widetilde{L}(\vec{\theta}^{(j)}),
	\label{eq:update}
\end{equation}
where $\eta$ is the learning rate, $j$ is the training step iteration and $\nabla_{\theta} \widetilde{L}(\cdot)$ is the estimate of the gradient. The size of the training batch determines the accuracy of the gradient. Such an update process based on an estimate of the gradient is commonly referred to as \ac{SGD}~\cite{goodfellow2016deep}.
Since all points of the constellation must appear multiple times in a training batch, the training batch size is specified in multiples of $M$.
In Fig.~\ref{fig:bsEntropy}, the convergence of the loss for different training batch sizes is shown.
A large training batch size leads to slower convergence but better final performance, and a small training batch size leads to faster convergence but worse final performance. The best trade-off between convergence speed, computation time and performance is achieved by starting the training process with smaller training batch size and increasing it after initial convergence. This is shown in Fig.~\ref{fig:bsEntropy}, where the green/dashed plot depicts a training run where the training batch size is increased after 100~iterations.
With a larger training batch size the statistics of the channel model is reflected more accurately. Changing the training batch size from small to large is reflected in a coarse-to-fine shift of the optimization process.
The size of the neural networks, number of layers and hidden units, are chosen depending on the order of the constellation. Other neural network hyperparameters, for both encoder and decoder, are given in Table~\ref{table:NNparams}.
\begin{table}[t]
    \caption{Hyperparameters of both encoder and decoder neural network.}
	\begin{tabularx}{\linewidth}{%
    @{}XX>{\centering\arraybackslash}X|@{}}
		Hyperparameter &  \\
	\hline
		\# Layers & 1-2 \\
		\# Hidden units per layer & 16-32  \\
		Learning rate & $\eta = 0.001$ \\
		Activation function & ReLU~\cite{nair2010rectified} \\
		Optimization method & Adam (SGD)~\cite{kingma2014adam}\\
		Training batch size & Adaptive (see Section \ref{subsec:AutoencoderModelTraining})
	\end{tabularx}	
	\label{table:NNparams}
\end{table}


\subsection{Mutual Information Estimation}\label{sec:neuralNetworkReceieverMI}
After training, only the learned constellation is used for transmission. The actual neural networks are neither implemented at the transmitter nor receiver. This is only feasible, since from the receiver standpoint both channel models appear Gaussian. This means, given the channel models, \ac{GN}-model or \ac{NLIN}-model, a receiver under a memoryless Gaussian auxiliary channel assumption is the \ac{ML} receiver~\cite{fehenberger2016probabilistic,eriksson2016impact}. Ultimately, the decoder neural network attempts to mimic this \ac{ML} receiver and the \ac{MI} could be estimated using the decoder neural network as in~\cite{li2018achievable}. Yet, the valid auxiliary channel assumption allows to estimate the \ac{MI} as described in~\cite{fehenberger2016probabilistic,eriksson2016impact}.
The encoder neural network is replaced by a look-up table without drawbacks, but during the training process both neural networks of the autoencoder model are required for their differentiability. With a dispersion free channel model~\cite{li2018achievable} or \ac{IM-DD} transmission~\cite{karanov2018end}, where a Gaussian auxiliary channel assumption does not hold, the trained decoder neural network or another receiver is required for detection.
\begin{figure}[t]
    \centering
    \includegraphics[width=\linewidth]{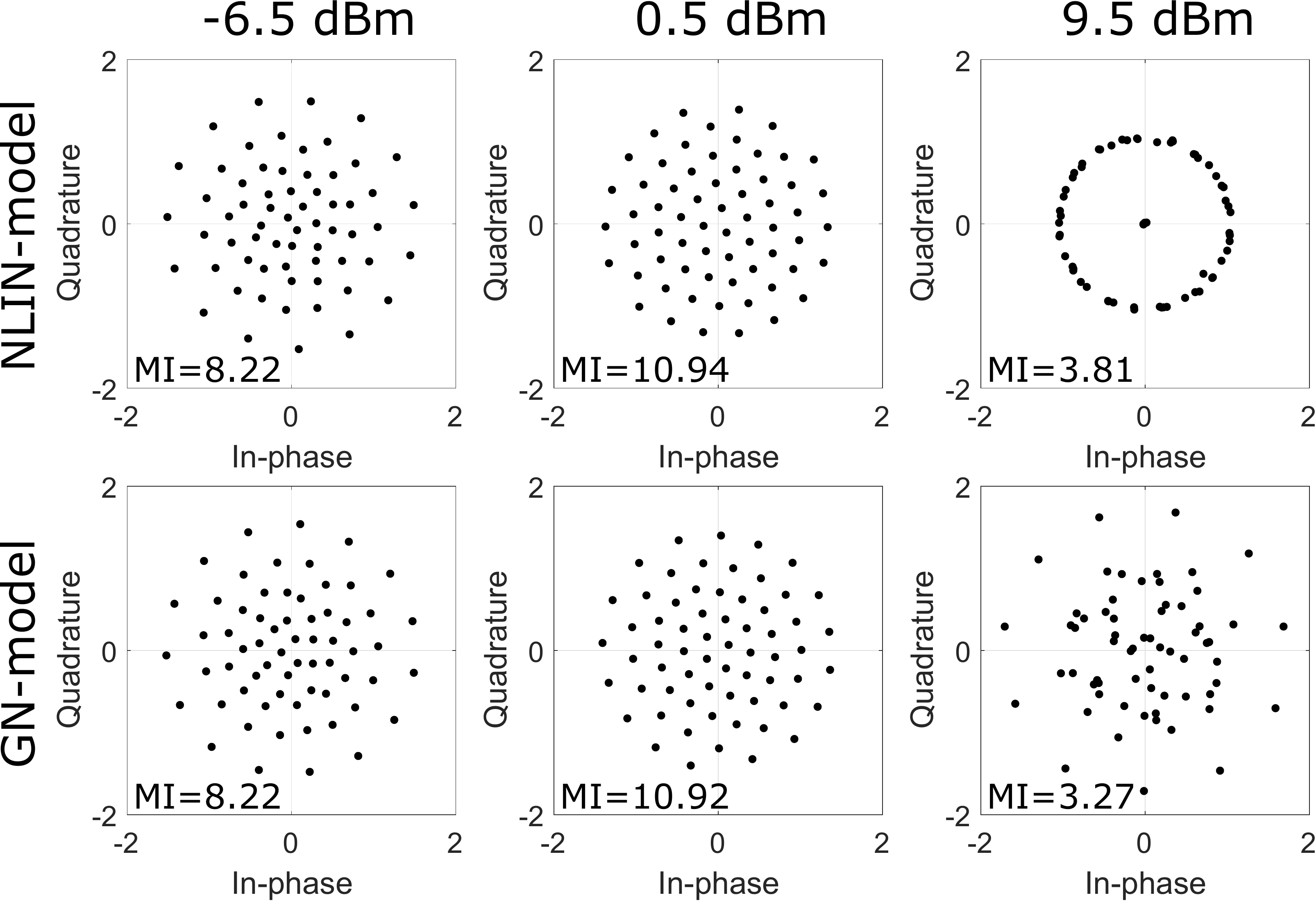}
    \caption{Constellations learned with \textbf{(top)} the \ac{NLIN}-model and \textbf{(bottom)} the \ac{GN}-model for $M$=64 and 1000~km (10 spans) transmission length at per channel launch power \textbf{(left to right)} $-$6.5~dBm, 0.5~dBm and 9.5~dBm.}
    \label{fig:constellations}
\end{figure}
\subsection{Joint optimization of constellation \& system parameters}
Composing and training of the autoencoder model is achieved with the machine learning framework TensorFlow~\cite{abadi2016tensorflow}. TensorFlow provides an interface to construct computation graphs. The neural networks and the channel model are implemented as such computation graphs. By exploiting the optimization capabilities of TensorFlow the neural networks are trained. However, the optimization is not limited to the parameters of the neural networks. Joint optimization of system parameters and the constellation is possible by simply extending $\vec{\theta}$, as long as the model stays differentiable w.r.t. $\vec{\theta}$. In Section \ref{sec:sim:jointOpt} joint optimization of the launch power and the constellation is studied with $\vec{\theta} = \{\vec{\theta}_f, \vec{\theta}_g, P_{\text{tx}} \}$. Thereby, also $P_{\text{tx}}$ is optimized in~(\ref{eq:update}).

\section{Geometrical Shaping for a \ac{WDM}-System via End-to-End Learning}
\label{sec:shapingViaEndToEndLearning}

\subsection{Learned Constellations}
A set of learned constellations optimized at increasing launch powers are shown in Fig.~\ref{fig:constellations}. The top and bottom row show constellations learned with the \ac{NLIN} and \ac{GN}-model, respectively.
At low powers~(left), the available \ac{SNR} is too low and the autoencoder found a constellation where the inner constellation points are randomly positioned. It represents one of many local minima of the loss function, that perform similar in \ac{MI}.
At optimal per channel launch power~(center), both constellations are very similar and also yield the same performance in \ac{MI}.
At high power levels~(right), the \ac{NLIN} is the primary impairment. All points in the \ac{NLIN}-model learned constellation either form a ring of uniform intensity or are located at the origin. This is because a ring-like constellation has minimized moments and minimizes the \ac{NLIN}. Again, there are many local minima, but now all are represented by ring-like constellations. The \ac{GN}-model learned constellations, trained under an \ac{AWGN} assumption independent of modulation dependent effects, lack this property. The autoencoder is bound to find one of many local minima like in the low power scenario.

\subsection{Simulation Setup}
\begin{table}[t]
    \caption{Simulation parameters of \ac{SSF} method simulations and \ac{NLIN}-model evaluations.}
	\begin{tabularx}{\linewidth}{%
    @{}XX>{\centering\arraybackslash}X|@{}}
		Simulation Parameter &  \\
	\hline
		\# symbols & $131072=\text{2}^{17}$ \\
		Symbol rate & 32 GHz \\
		Oversampling & 32 \\
		Channel spacing & 50 GHz \\
		\# Channels & 5 \\
		\# Polarisation & 2 \\
		Pulse shaping & root-raised-cosine \\
		Roll-off factor & SSF: 0.05 \\
		                & NLIN-model: 0 (Nyquist) \\
        Span length & 100 km \\
		Nonlinear coefficient & 1.3 $(\text{W km})^{-1}$\\
		Dispersion parameter & 16.48 $\text{ps}/(\text{nm km})$\\
		Attenuation & 0.2 dB/km \\
		\ac{EDFA} noise figure & 5 dB \\
		Stepsize & 0.1 km \\
	\end{tabularx}
	\label{table:SSF}
\end{table}

A \ac{WDM} communication system was simulated using the \ac{SSF} method. The parameters of the transmission are given in Table~\ref{table:SSF}. The simulated transmitter included digital pulse shaping and ideal optical I/Q modulation. All \ac{WDM} channels were modulated with independent and identically distributed drawn symbol sequences from the same constellation. The channel included multiple spans with lumped amplification, where \ac{EDFA}s introduced \ac{ASE} noise. The receiver included an optical filter for the center channel, digital compensation of chromatic dispersion and matched filtering. Besides numerical simulations, the \ac{NLIN}-model was also used for performance evaluations of the constellations. This allowed a comparison of the autoencoder learned constellations, \ac{QAM} constellations and \ac{IPM}-based geometrically shaped constellations.

\begin{figure}[t]
    \centering
    \begin{minipage}{0.5\textwidth}
        \centering
        \includegraphics[width=0.8\linewidth]{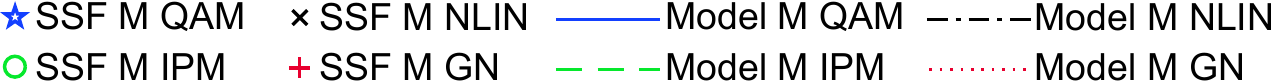}
        \vspace{4pt}
    \end{minipage}
    \begin{minipage}{0.25\textwidth}
        \centering
        \includegraphics[width=\linewidth]{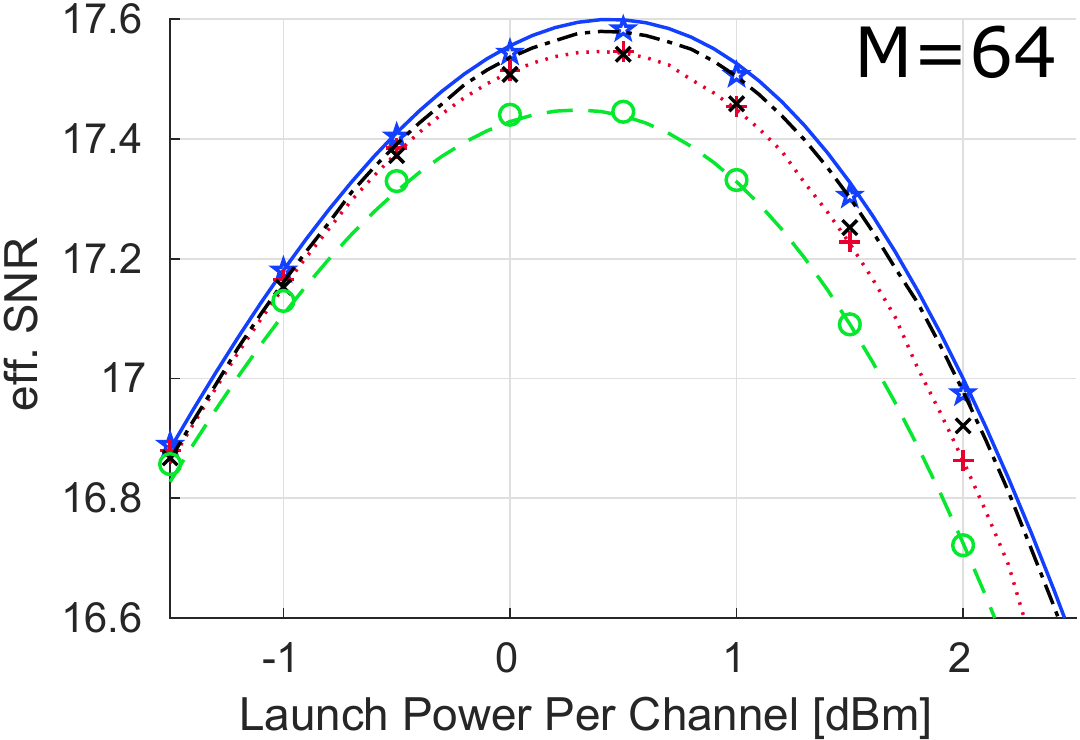}
    \end{minipage}%
    \begin{minipage}{0.25\textwidth}
        \centering
        \includegraphics[width=\linewidth]{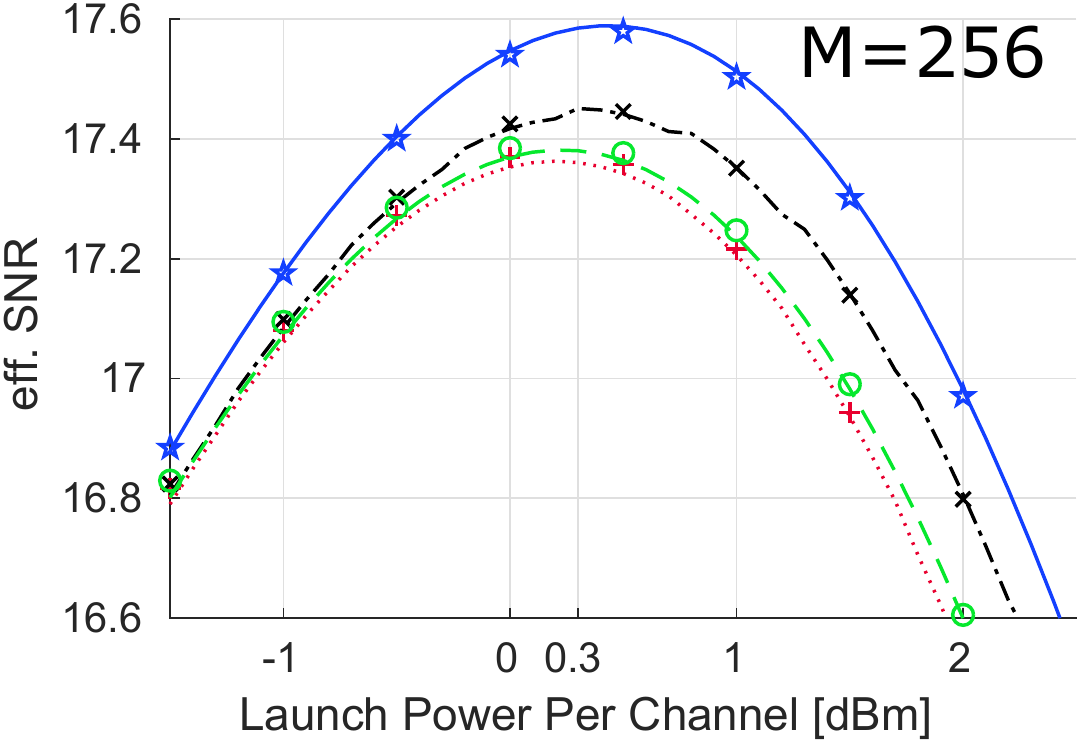}
    \end{minipage}
    \begin{minipage}{0.25\textwidth}
        \centering
        \includegraphics[width=\linewidth]{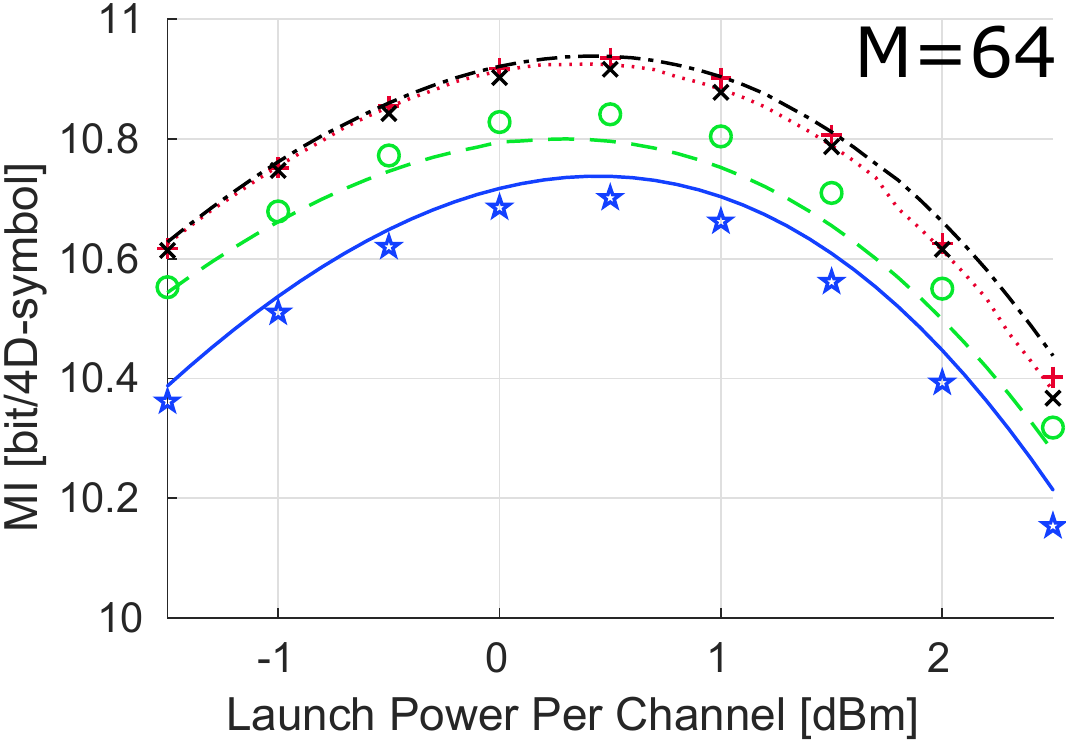}
    \end{minipage}%
    \begin{minipage}{0.25\textwidth}
        \centering
        \includegraphics[width=\linewidth]{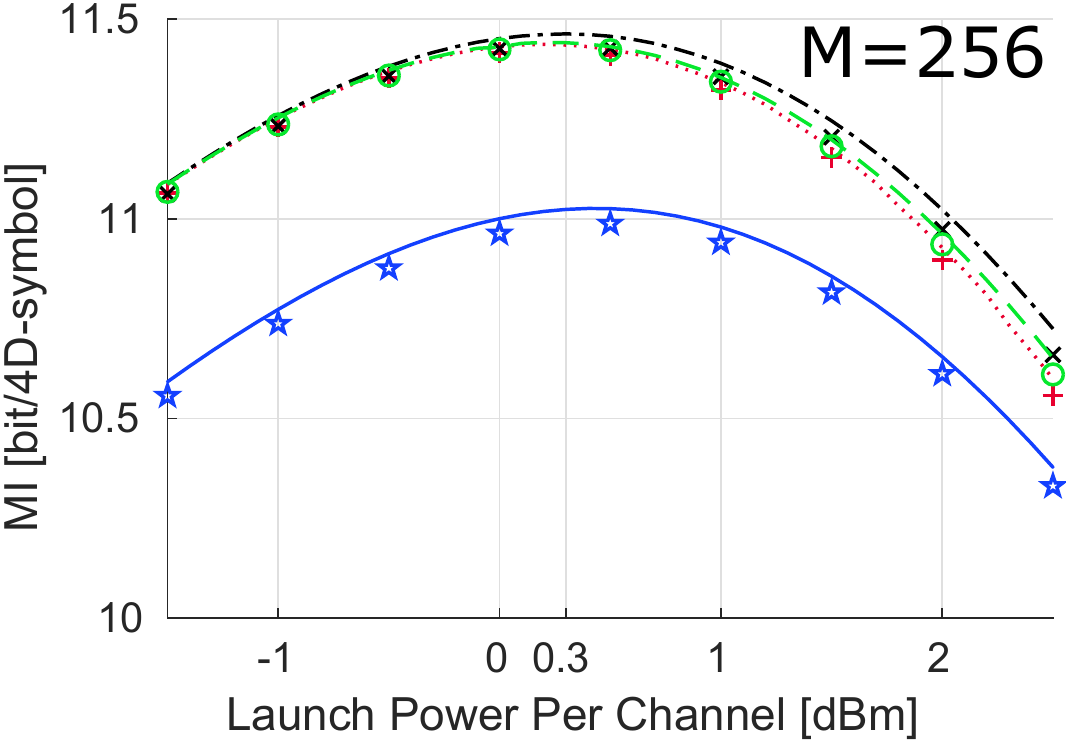}
    \end{minipage}    
    \caption{Performance in \textbf{(top)} effective \ac{SNR} and \textbf{(bottom)} \ac{MI} with respect to launch power after 1000~km transmission (10 spans) for \textbf{(left)} $M$=64 and \textbf{(right)} $M$=256. Plots denoted as "$M$~\ac{GN}" and "$M$~\ac{NLIN}" indicate that the constellation was learned using the \ac{GN}-model and \ac{NLIN}-model, respectively. Lines depict performance evaluations using the \ac{NLIN}-model and markers using the \ac{SSF} method.}
    \label{fig:simulationResults}
\end{figure}

\begin{figure}[t]
	\centering
	\includegraphics[width=\linewidth]{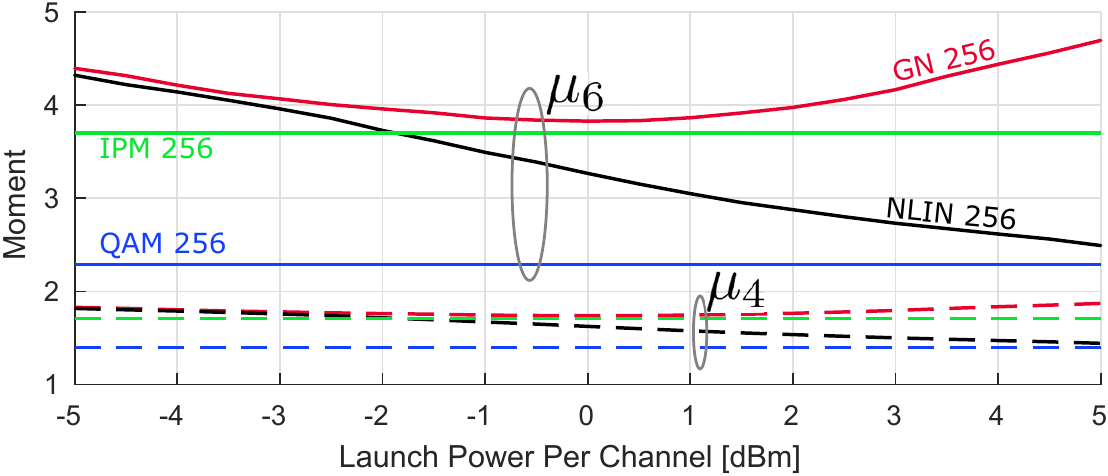}
    \caption{4th and 6th order moment ($\mu_4$ and $\mu_6$) of optimized constellations with $M$=256 with respect to the per channel launch power and 1000~km transmission distance.}
    \label{fig:ConstMoment}
\end{figure}

\begin{figure}[t]
	\centering
	\includegraphics[width=\linewidth]{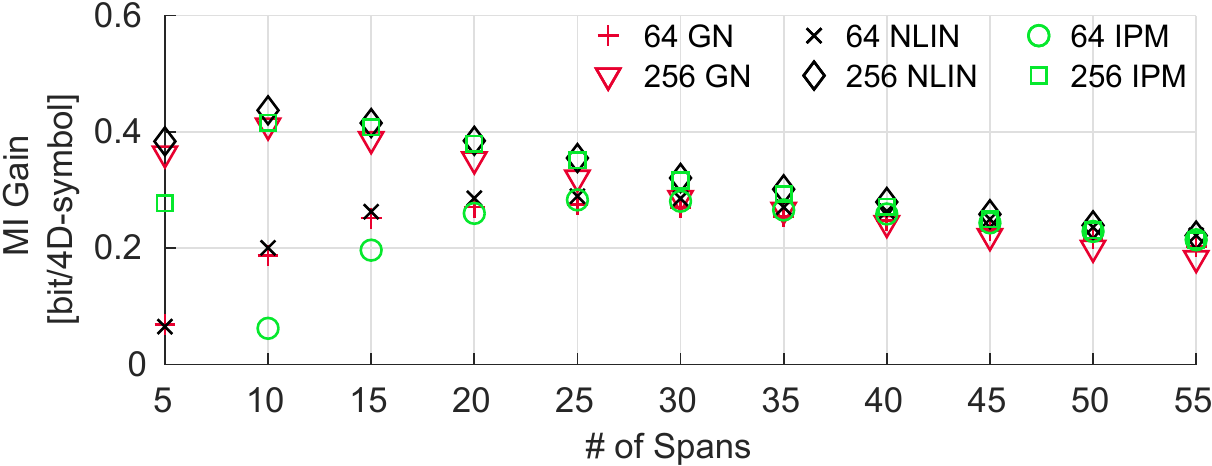}
    \caption{Gain compared to the standard $M$-\ac{QAM} constellations at the respective optimal launch power with respect to the number of spans estimated with the \ac{NLIN}-model.}
    \label{fig:DeltaMiSpan}
\end{figure}

\begin{figure}[t]
    \centering
    \begin{minipage}{0.25\textwidth}
        \centering
        \includegraphics[width=\linewidth]{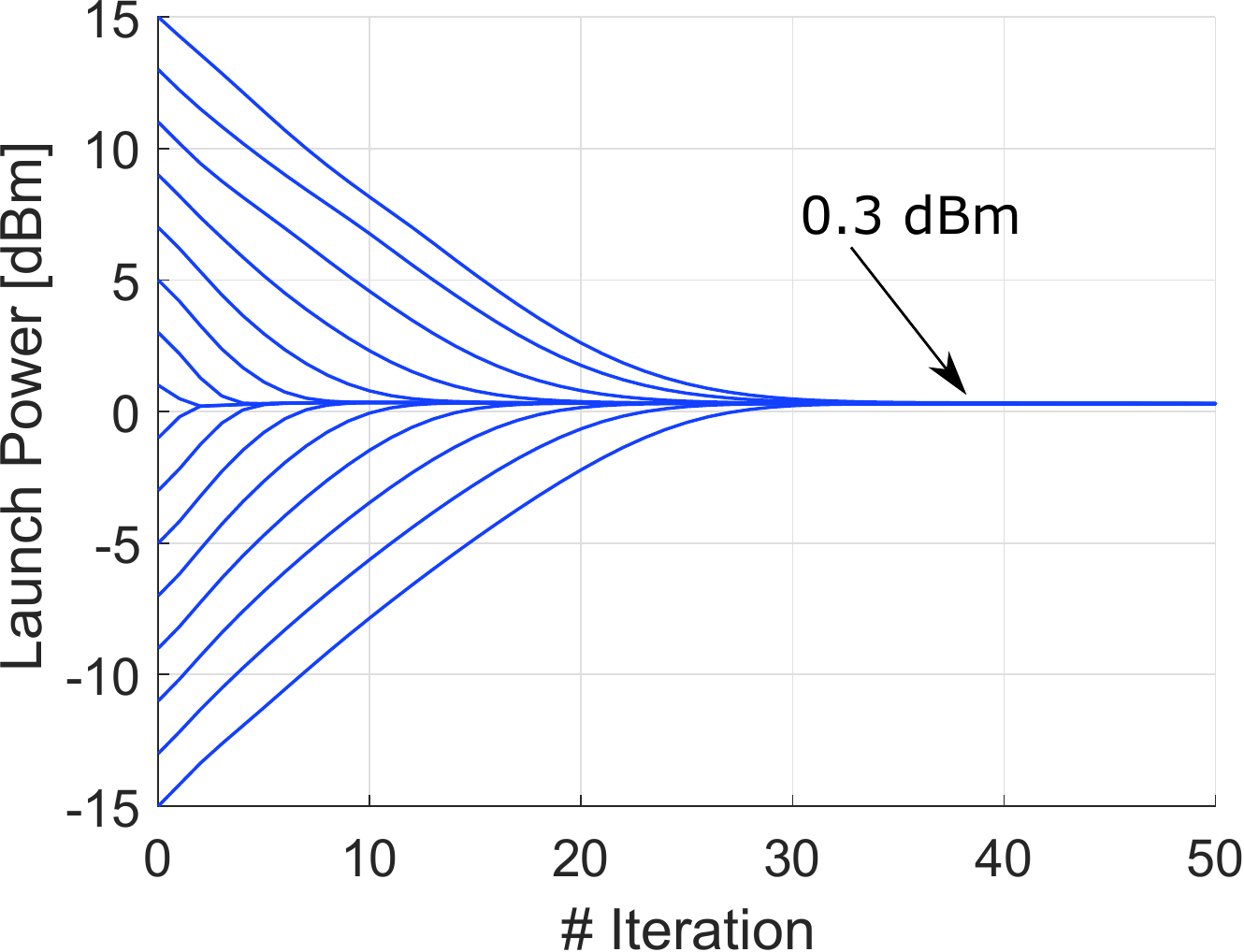}
    \end{minipage}
    \begin{minipage}{0.25\textwidth}
        \centering
        \includegraphics[width=\linewidth]{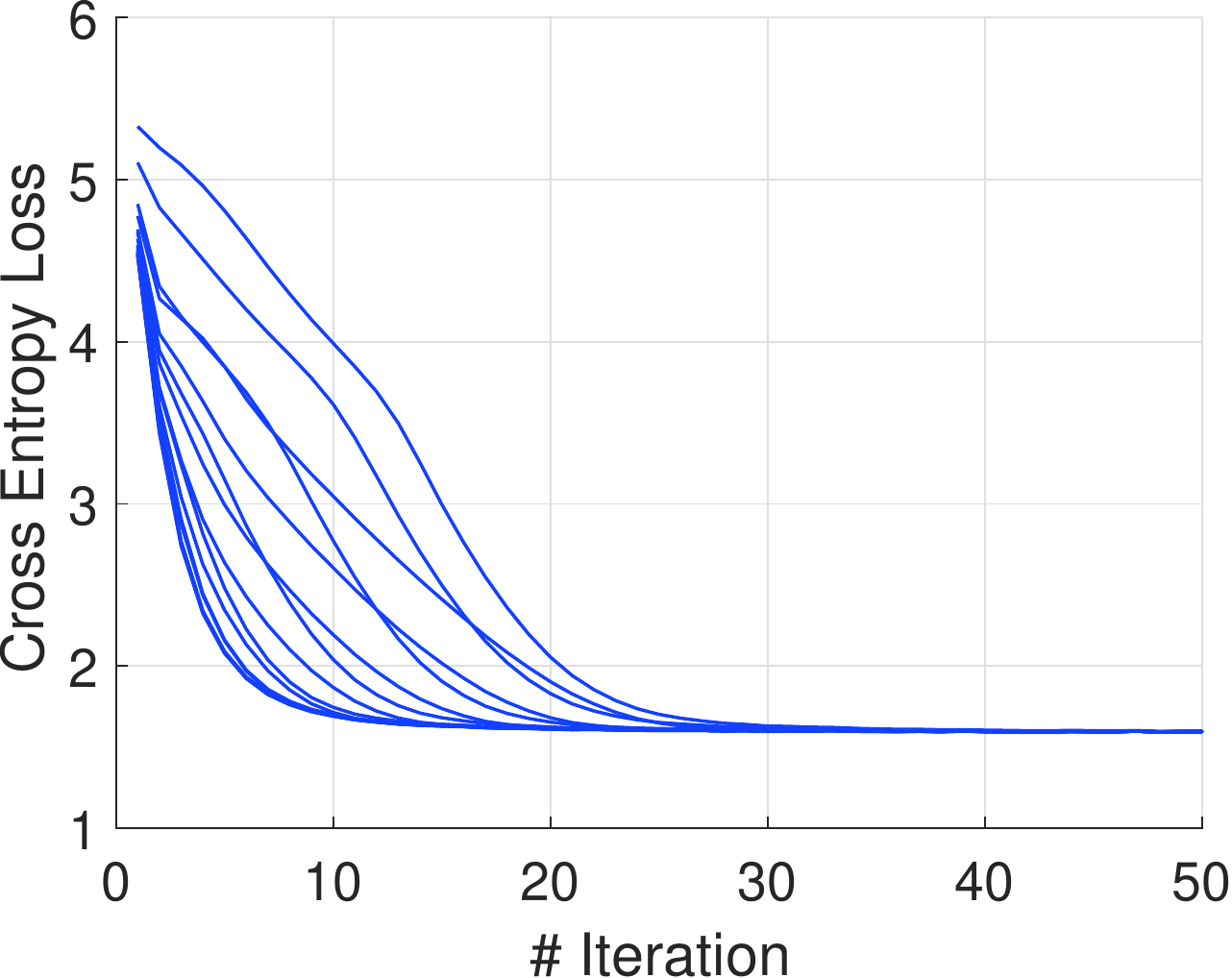}
    \end{minipage}
    \caption{\textbf{(left)} Convergence of the per channel launch power starting from different initial values, and \textbf{(right)} respective cross entropy loss, when jointly learning the constellation shape and launch power. The constellation is learned using the \ac{NLIN}-model for $M$=256.}
    \label{fig:pOpt}
\end{figure}
\begin{figure*}[t]
    \centering
    \includegraphics[width=\linewidth]{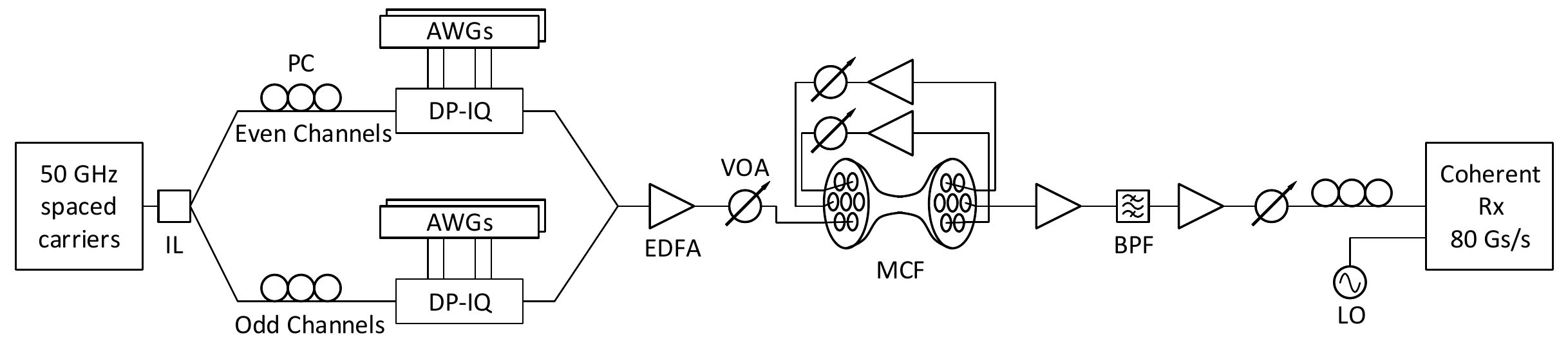}
    \caption{Experimental setup. Abbr.: (IL) Interleaver, (PC) Polarization Controller, (AWG) Arbitrary Waveform Generator, (DP-IQ) Dual-Polarization IQ Modulator, (EDFA) Erbium Doped Fiber Amplifier, (VOA) Variable Optical Attenuator, (MCF) Multicore Fiber (used to emulate 3 identical single-mode fiber spans), (BPF) Band-pass Filter, (LO) Local Oscillator}
    \label{fig:experimentalSetup}
\end{figure*}

\subsection{Simulation Results}
\label{sec:sim:results}
The constellations were evaluated using the estimated received \ac{SNR} and the \ac{MI}, which constitutes the maximum \ac{AIR} under an auxiliary channel assumption~\cite{fehenberger2016probabilistic}. The simulation results for a 1000~km transmission (10~spans) are shown in Fig.~\ref{fig:simulationResults}.
The autoencoder constellations show improved performance in \ac{MI} compared to standard \ac{QAM} constellations and improved effective \ac{SNR} compared to \ac{IPM}-based constellations. \ac{QAM} constellations have the best effective \ac{SNR} due to their low high-order moments, but are clearly not shaped for gain in \ac{MI}.
Further, the optimal launch power for constellations trained on the \ac{NLIN}-model is shifted towards higher powers compared to those of the \ac{GN}-model and \ac{IPM}-based constellations.
In Fig.~\ref{fig:simulationResults}, at higher powers the gain with respect to standard \ac{QAM} constellations is larger for the \ac{NLIN}-model obtained constellations, and in Fig.~\ref{fig:ConstMoment}, the constellations learned with the \ac{NLIN}-model have a negative slope in moment.
This shows that the autoencoder wrapping the \ac{NLIN}-model has learned constellations with smaller higher-order moments and smaller nonlinear impairment.
At the optimal launch power and transmission distances of 2500~km to 5500~km, the difference between \ac{NLIN}-model, \ac{GN}-model and \ac{IPM}-based constellations was marginal, as shown in Fig.~\ref{fig:DeltaMiSpan}.
However, at lower transmission distance the autoencoder learned constellations outperformed the \ac{IPM}-based constellations.
For a transmission distance of 1000~km (10~spans) at the respective optimal launch power, the learned constellations yielded 0.13~bit/4D and 0.02~bit/4D higher \ac{MI} than \ac{IPM}-based constellations evaluated with the \ac{NLIN}-model for $M$=64 and $M$=256, respectively. Evaluated with the \ac{SSF} method, the learned constellations yielded 0.08~bit/4D and 0.00~bit/4D improved performance in \ac{MI} compared to \ac{IPM}-based constellations for $M$=64 and $M$=256, respectively.

\subsection{Joint optimization of constellation \& launch power}
\label{sec:sim:jointOpt}
The simulation results of Section \ref{sec:sim:results} show that the optimal per channel launch power of the $M$=256 \ac{NLIN}-model learned constellation was 0.3~dBm. The optima can also be estimated within the training process, by including the launch power to the trainable parameters of the autoencoder model, $\vec{\theta} = \{\vec{\theta}_f, \vec{\theta}_g, P_{\text{tx}} \}$. The training process is started with an initial guess for its parameters. In Fig.~\ref{fig:pOpt}~(left) it is shown, that the per channel launch power estimate always converged towards the optimal value of 0.3~dBm, even when starting the training process with different initial values. Further, the performances of all training runs converged in terms of cross entropy loss independent from the initial value, Fig.~\ref{fig:pOpt}~(right).

\subsection{Experimental Demonstration}
\label{sec:exp:demo}
The experimental setup is shown in Fig.~\ref{fig:experimentalSetup}. The 5 channel \ac{WDM} system was centered at 1546.5~nm with a 50~GHz channel spacing with the carrier lines extracted from a 25~GHz comb source using one port of a 25~GHz/50~GHz optical interleaver. Four \ac{AWG} with a baud rate of 24.5~GHz were used to modulate two electrically decorrelated signals onto the odd and even channels, generated in a second 50~GHz/100~GHz optical interleaver, (IL in Fig.~\ref{fig:experimentalSetup}). The fiber channel was comprised of 1 to 3 fiber spans with lumped amplification by \ac{EDFA}s.
In the absence of identical spans, we chose to use 3 far separated outer cores of a multicore fiber. The multicore fiber was of 53.5~km length and the crosstalk between cores was measured to be negligible.
The launch power in to each span was adjusted by \ac{VOA} placed after in-line \ac{EDFA}s. 
Hence, we note that scanning the fiber launch power would also change the input power to post-span EDFAs that may slightly impact the noise performance.
At the receiver, the center channel was optically filtered and passed on to the heterodyne coherent receiver. The \ac{DSP} was performed offline. It involved resampling, digital chromatic dispersion compensation, and data aided algorithms for both the estimation of the frequency offset and the filter taps of the \ac{LMS} dual-polarization equalization. The equalization integrated a blind phase search. The data aided algorithms are further discussed in Section~\ref{sec:dataaid}.
The autoencoder model was enhanced to include transmitter imperfections in terms of an additional \ac{AWGN} source in between the normalization and the channel. The back-to-back \ac{SNR} was measured for \ac{QAM} constellations and yielded 21.87~dB.
We note that it was challenging to reliably measure some experimental features such as transmitter imperfections and varying \ac{EDFA} noise performance and accurately represent them in the model. Hence, it is likely that some discrepancy between the experiment and modeled link occurred, as further discussed in Section~\ref{sec:missmatch}.

\subsection{Experimental Results}
\begin{figure*}[t]
    \centering
    \begin{minipage}{1\textwidth}
        \centering
        \includegraphics[width=0.5\linewidth]{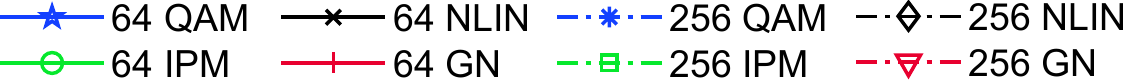}
        \vspace{4pt}
    \end{minipage}    
    \begin{minipage}{0.33\textwidth}
        \centering
        1 Span
    \end{minipage}%
    \begin{minipage}{0.33\textwidth}
        \centering
        2 Spans
    \end{minipage}%
    \begin{minipage}{0.33\textwidth}
        \centering
        3 Spans
    \end{minipage}
    \begin{minipage}{0.33\textwidth}
        \centering
        \includegraphics[width=\linewidth]{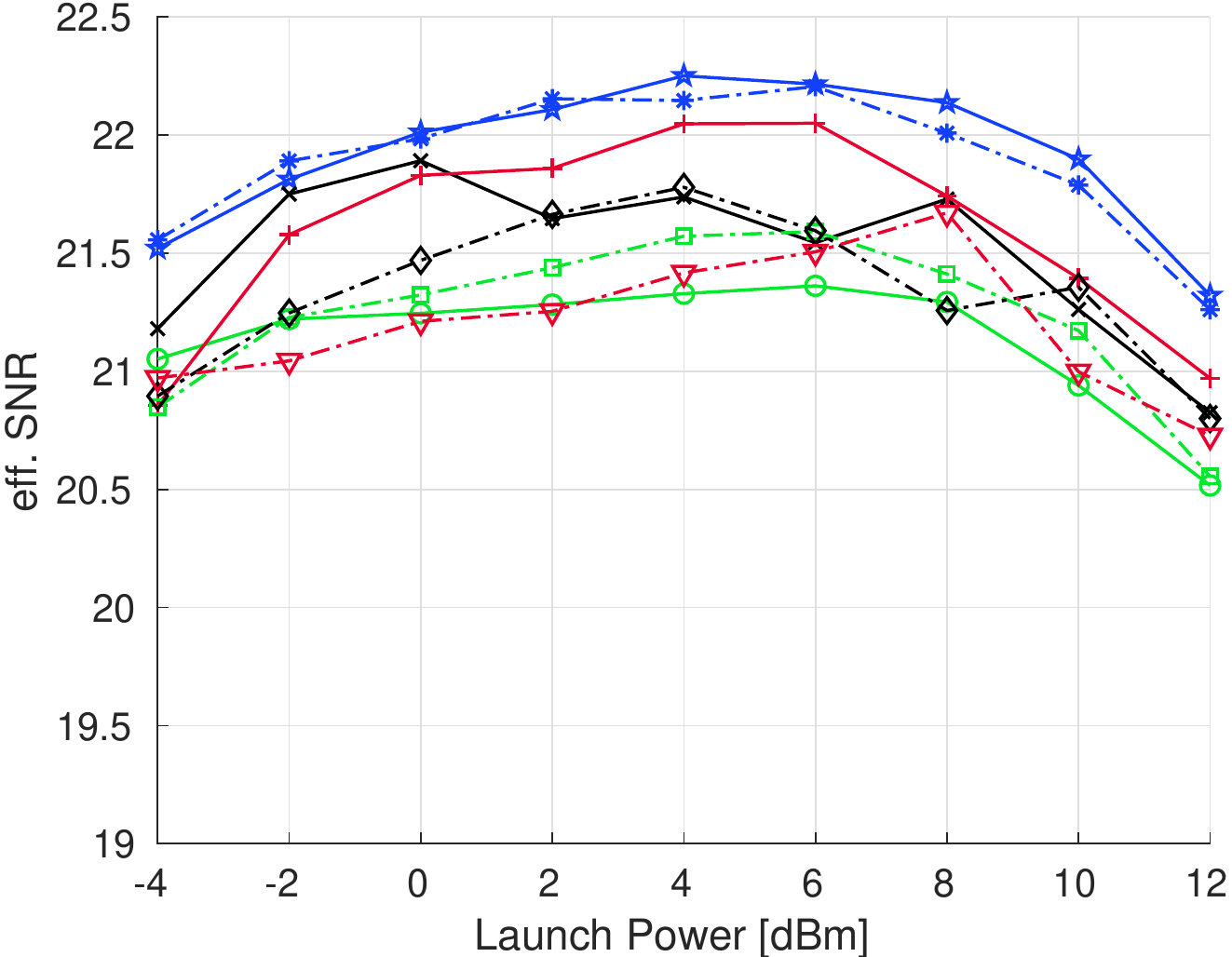}
    \end{minipage}%
    \begin{minipage}{0.33\textwidth}
        \centering
        \includegraphics[width=\linewidth]{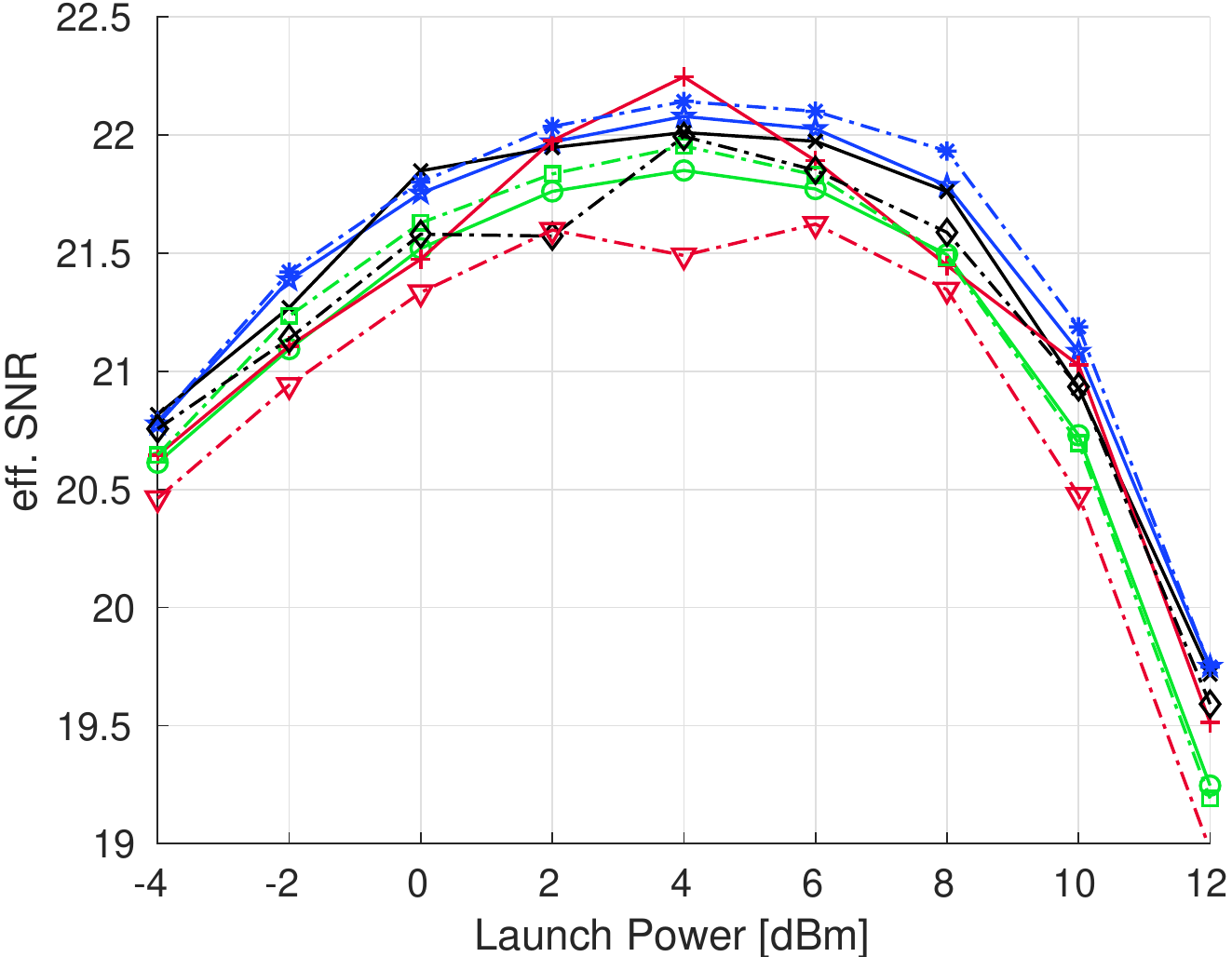}
    \end{minipage}%
    \begin{minipage}{0.33\textwidth}
        \centering
        \includegraphics[width=\linewidth]{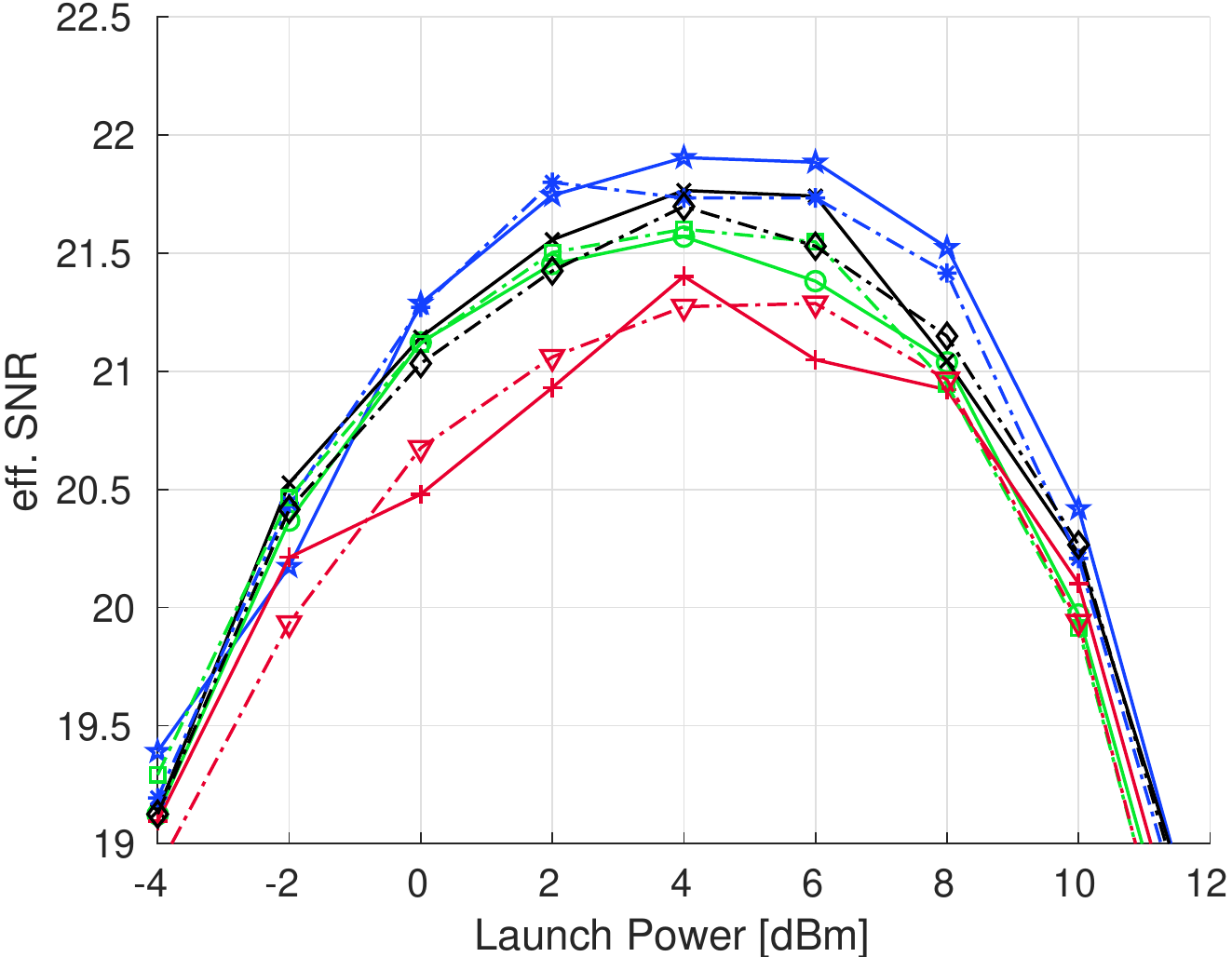}
    \end{minipage}
    \begin{minipage}{0.33\textwidth}
        \centering
        \includegraphics[width=\linewidth]{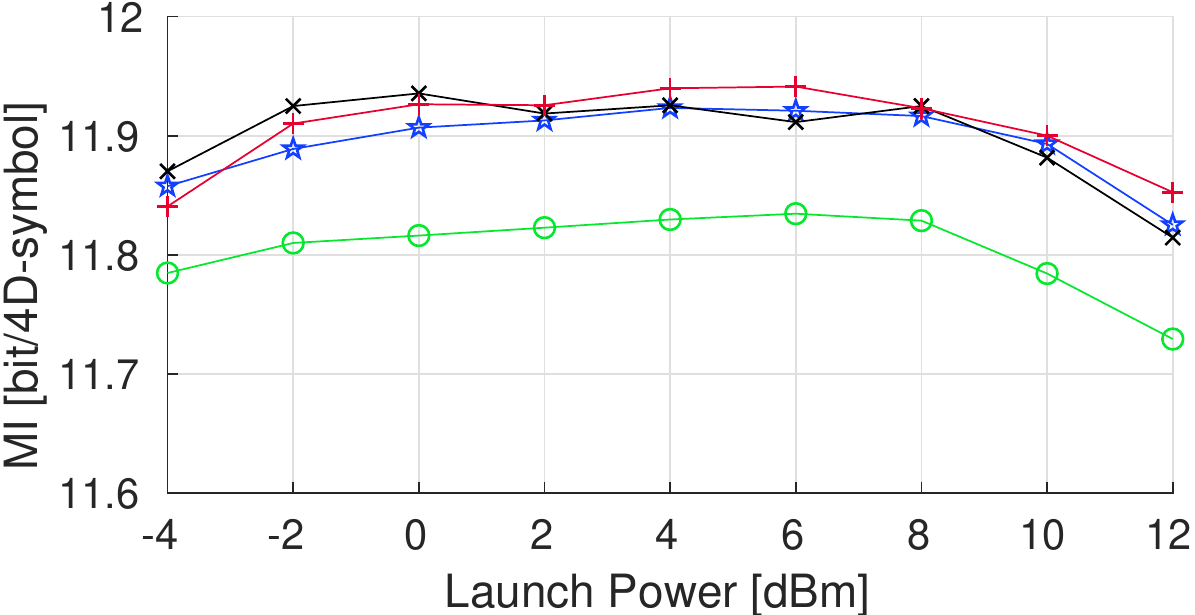}
    \end{minipage}%
    \begin{minipage}{0.33\textwidth}
        \centering
        \includegraphics[width=\linewidth]{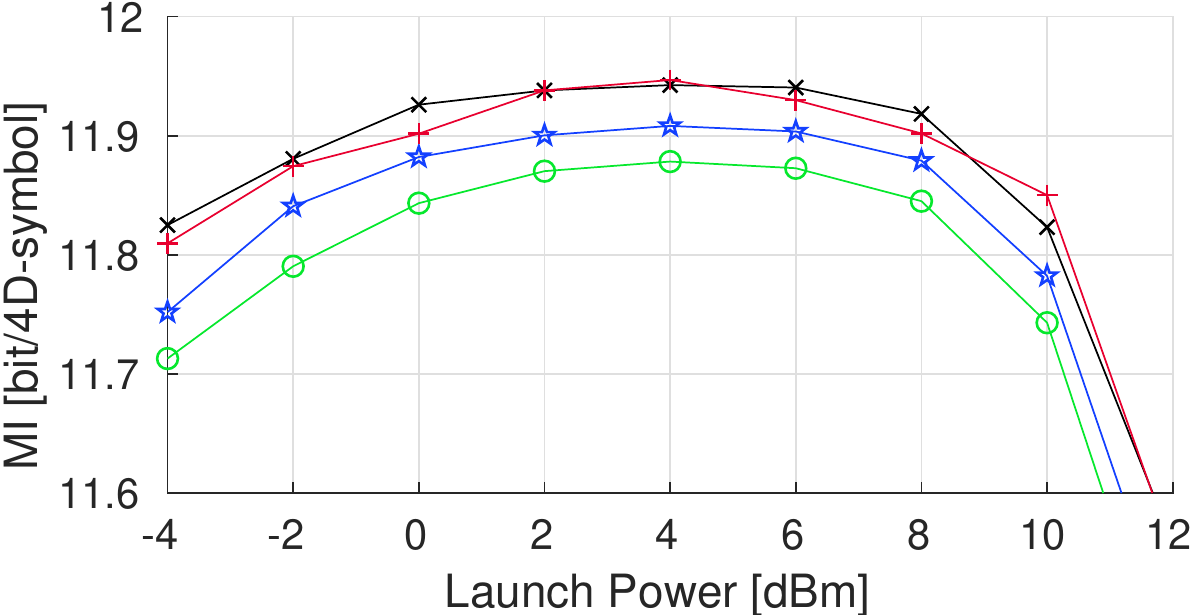}
    \end{minipage}%
    \begin{minipage}{0.33\textwidth}
        \centering
        \includegraphics[width=\linewidth]{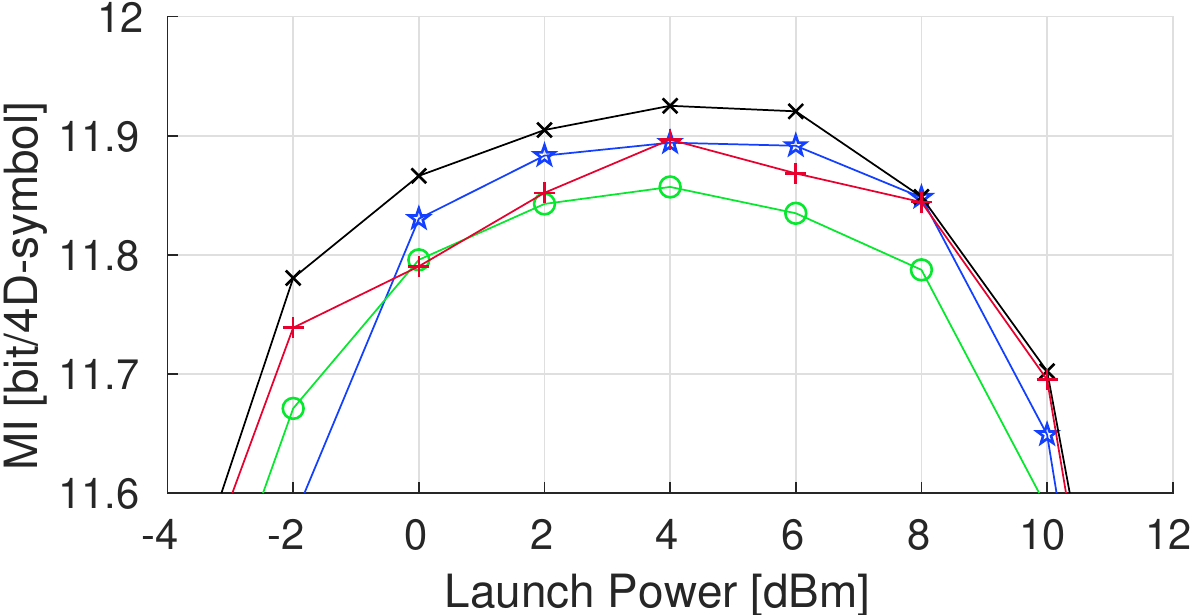}
    \end{minipage}
    \begin{minipage}{0.33\textwidth}
        \centering
        \includegraphics[width=\linewidth]{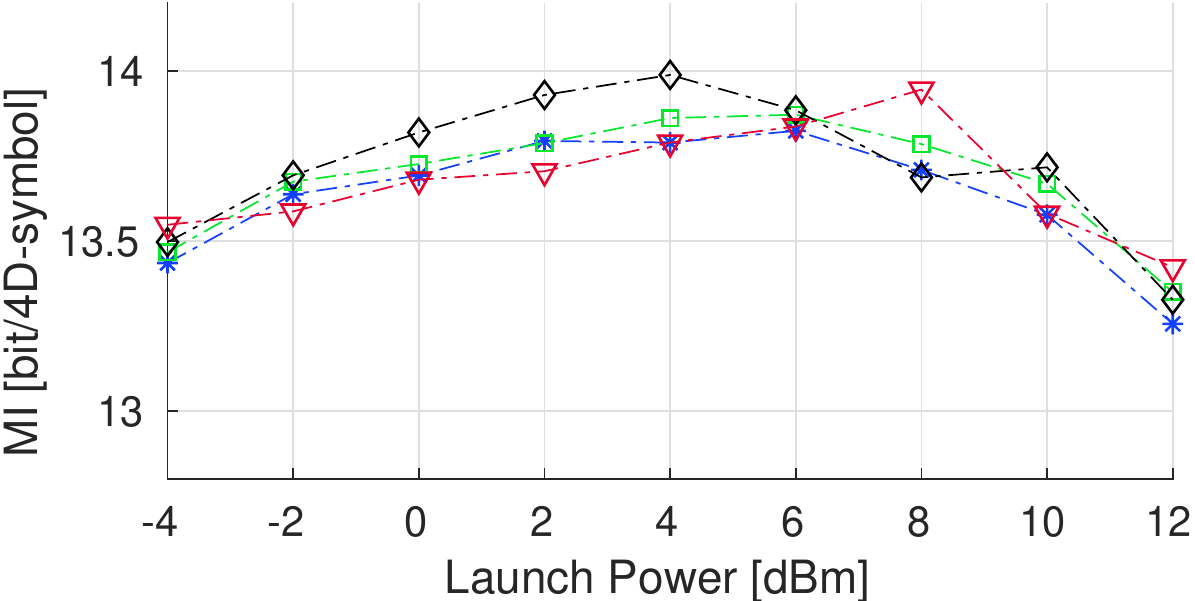}
    \end{minipage}%
    \begin{minipage}{0.33\textwidth}
        \centering
        \includegraphics[width=\linewidth]{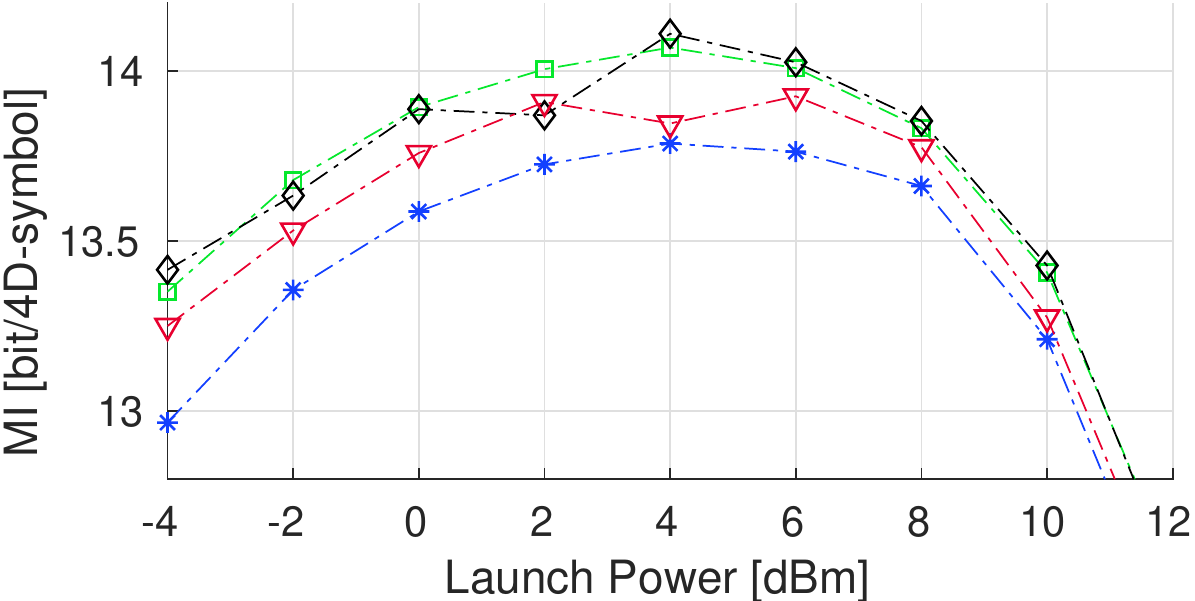}
    \end{minipage}%
    \begin{minipage}{0.33\textwidth}
        \centering
        \includegraphics[width=\linewidth]{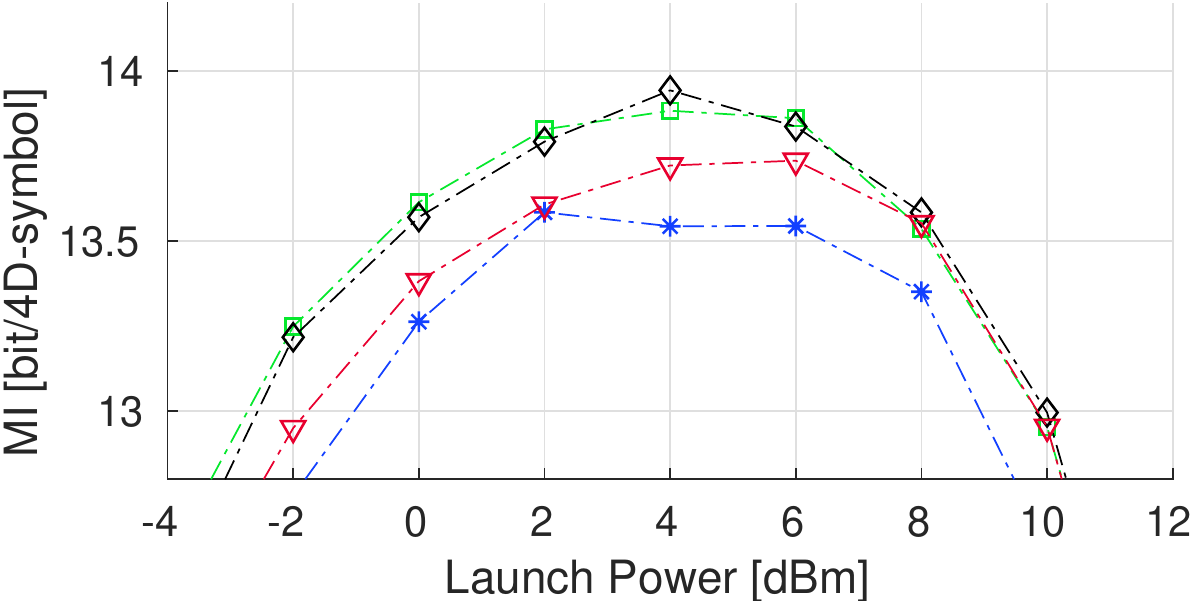}
    \end{minipage}
    \caption{Experimental Results. \textbf{(top)}~Effective \ac{SNR} in respect to the launch power for $M$=64 and $M$=256 constellations. \textbf{(mid)}~\ac{MI} in respect to the launch power for $M$=64 constellations. \textbf{(bottom)}~\ac{MI} in respect to the launch power for $M$=256 constellations. 
    \textbf{(left)}, \textbf{(center)} and \textbf{(right)} depict the results of one, two and three span transmissions, respectively.}
    \label{fig:experimentalResults}
\end{figure*}

The experimental demonstration confirmed the trend of the simulation results, but some discrepancies were observed and attributed to a mismatch between simulation parameters and those used in the experiment. The constellations were evaluated using the estimated received \ac{SNR} and the \ac{MI}, which constitutes the maximum \ac{AIR} under an auxiliary channel assumption~\cite{fehenberger2016probabilistic}.
In Fig.~\ref{fig:experimentalResults}, the performances are compared to \ac{QAM} and \ac{IPM}-based constellations. In general, the learned constellations outperformed the \ac{IPM}-based and \ac{QAM} constellations.
However, we observe that with more spans, the discrepancy between the experiment and modeled link increases.
Over one span, Fig.~\ref{fig:experimentalResults}~(left), the learned constellations outperformed \ac{QAM} and \ac{IPM}-based constellations. For $M$=64 the \ac{IPM}-based constellations performed worst.
This is due to the relatively high \ac{SNR} and an \ac{AIR} close to the upper bound of $2 \cdot \log_2 M$, at which irregular constellations suffer a penalty w.r.t. regular \ac{QAM}. 
Over two spans and $M$=64, Fig.~\ref{fig:experimentalResults}~(center), \ac{IPM}-based constellations yielded improved performances. Over two spans and for $M$=256, they outperformed the \ac{GN}-model learned constellations and matched the performance of the \ac{NLIN}-model learned constellations. This indicates, that the \ac{NLIN}-model learned constellations failed to mitigate nonlinear effects due to the model discrepancy. Similar findings to the transmissions over two spans were found over three spans, Fig.~\ref{fig:experimentalResults}~(right).
Collectively, the learned constellations, either learned with the \ac{NLIN}-model or the \ac{GN}-model,
yielded improved performance compared to either \ac{QAM} or \ac{IPM}-based constellations at their respective optimal launch power. For $M$=64 the improved performance was 0.02~bit/4D, 0.04~bit/4D and 0.03~bit/4D in \ac{MI} over 1, 2 and 3 spans transmission, respectively. For $M$=256 the improved performance was 0.12~bit/4D, 0.04~bit/4D and 0.06~bit/4D in \ac{MI} over 1, 2 and 3 spans transmission, respectively.

\section{Discussion}
\label{sec:discussion}
\subsection{Mitigation of Nonlinear Effects}
The \ac{NLIN}-model describes a relationship between the launch power, the high-order moments of the constellation and signal degrading nonlinear effects. Thus, there exist geometric shaped constellations that minimize nonlinear effects by minimizing their high-order moments. In contrast, constellations robust to Gaussian noise are Gaussian shaped, and hence have large high-order moments. This means, the minimization only improves the system performance if the relative contributions of the modulation dependent nonlinear effects are significant compared to those of the \ac{ASE} noise. If the relative contributions to the overall signal degradation of both noise sources are similar, an optimal constellation must be optimized for a trade-off between the two. The autoencoder model is able to capture these channel characteristics, as shown by the higher robustness of the \ac{NLIN}-model learned constellations. At higher powers these learned constellations yield a gain in \ac{MI} and \ac{SNR} compared to \ac{IPM}-based geometrically shaped constellations and constellations learned with the \ac{GN}-model. Further, in Fig.~\ref{fig:ConstMoment}, the high-order moments of the more robust constellations decline with larger launch power.
This shows that nonlinear effects are mitigated.
At larger transmission distance, Fig.~\ref{fig:DeltaMiSpan}, modulation format dependent nonlinear effects become less dominant and therefore also the gains of the constellations learned with the \ac{NLIN}-model are reduced.

\subsection{Model Discrepancy}\label{sec:missmatch}
The experimental results show that the learned constellations match the performance of \ac{IPM}-based geometrically shaped constellations. However, it is not obvious that a constellation learned with the \ac{NLIN}-model also mitigates nonlinear effects experimentally. 
For this, good matching of the channel model parameters to those of the experimental setup is required.
Unknown fiber parameters, imperfect components and device operating points make this very challenging. In particular, transmitter imperfections due to digital-to-analog conversion are hard to capture in the model, since they can be specific to a particular constellation. For the experimental demonstration in Section~\ref{sec:exp:demo}, the \ac{SNR} at the transmitter was measured for a \ac{QAM} constellation and included as \ac{AWGN} source into the autoencoder model. Although, extending the autoencoder model to include further transmitter/receiver components is desirable, it also involves adding more unknowns. The sensitivity of the presented method to such uncertainties must be further studied. Methods introducing reinforcement learning~\cite{aoudia2018end}, or adversarial networks~\cite{o2018physical} can potentially avoid the modeling discrepancy and learn in direct interaction with the real system.

\subsection{Digital Signal Processing Drawbacks}
\label{sec:dataaid}
Standard off-the-shelf signal recovery algorithms often exploit symmetries inherent in \ac{QAM} constellations. This is a typical issue for non-standard \ac{QAM} constellations, including the ones designed in this work. In such cases, pilot-aided, constellation independent \ac{DSP} algorithms are typically required~\cite{millar2016design,yankov2017experimental}. In order to receive the learned constellations, the frequency offset estimation implements an exhaustive search based on temporal correlations. It correlates the received signal with many different frequency shifted versions of the transmitted signal until the matching frequency offset is found. The adaptive \ac{LMS} equalizer taps are estimated with full knowledge of the transmitted sequence. The blind phase search is integrated within the equalizer, but has no knowledge of the data and searches across all angles instead of only angles of one quadrant.
Other constellation shaping methods for the fiber optic channel which constrain the solution space to square constellations preserve the required symmetries~\cite{sillekens2018experimental}.

\section{Conclusion}
\label{sec:conclusion}
A new method for geometric shaping in fiber optic communication systems is shown, by leveraging an unsupervised machine learning algorithm, the autoencoder. 
With a differentiable channel model including modulation  dependent  nonlinear  effects, the learning algorithm yields a constellation mitigating these, with gains up to 0.13~bit/4D in simulation and up to 0.12~bit/4D experimentally. The machine learning optimization method is independent of the embedded channel model and allows joint optimization of system parameters. The experimental results show improved performances but also highlight challenges regarding matching the model parameters to the real system and limitations of conventional \ac{DSP} for the learned constellations.

\section*{Acknowledgment}
We thank Dr. Tobias Fehenberger for discussions on the used channel models. This work was financially supported by Keysight Technologies (Germany, B\"oblingen) and by the European Research Council through the ERC-CoG FRECOM project (grant agreementno. 771878).

\ifCLASSOPTIONcaptionsoff
  \newpage
\fi

\footnotesize












\acrodef{NLSE}[NLSE]{Nonlinear Schr\"odinger equation}
\acrodef{AWGN}[AWGN]{additive white Gaussian noise}
\acrodef{SNR}[SNR]{signal-to-noise ratio}
\acrodef{GN}[GN]{Gaussian noise}
\acrodef{NLIN}[NLIN]{nonlinear interference noise}
\acrodef{MI}[MI]{mutual information}
\acrodef{SSF}[SSF]{split-step Fourier}
\acrodef{ASE}[ASE]{amplified spontaneous emission}
\acrodef{IQ}[IQ]{in-phase and quadrature}
\acrodef{WDM}[WDM]{wavelength-division multiplexing}
\acrodef{QAM}[QAM]{quadrature amplitude modulation}
\acrodef{EN}[EN]{encoder network}
\acrodef{DN}[DN]{decoder network}
\acrodef{NN}[NN]{neural network}
\acrodef{IPM}[IPM]{iterative polar modulation}
\acrodef{EDFA}[EDFA]{erbium doped fiber amplifiers}
\acrodef{RTN}[RTN]{radio transformer networks}
\acrodef{LMS}[LMS]{least mean squares}
\acrodef{AWG}[AWG]{arbitrary waveform generators}
\acrodef{SSMF}[SSMF]{standard single-mode fiber}
\acrodef{VOA}[VOA]{variable optical attenuators}
\acrodef{DSP}[DSP]{digital signal processing}
\acrodef{SGD}[SGD]{stochastic gradient descent}
\acrodef{EGN}[EGN]{extended Gaussian noise}
\acrodef{IM-DD}[IM-DD]{intensity modulation direct detection}
\acrodef{ML}[ML]{maximum likelihood}
\acrodef{AIR}[AIR]{achievable information rate}

\end{document}